\documentclass{aastex62}
\usepackage{amsmath}
\usepackage{amsfonts}
\usepackage{amssymb}
\usepackage{multirow}
\usepackage{graphicx}
\usepackage{booktabs}

\def \apjs {{\rm {ApJS}}}

\def \aap {{\rm {A\&A}}}

\def \apjl{\rm {ApJL}}

\def \deg{$^\circ$}

\date{\today}
\received{xxx}
\revised{xxx}
\accepted{xxx}
\submitjournal{xx}

\shorttitle{Search for nearby Earth analogs}
\shortauthors{Feng et al.}

\begin{document}
\title{Search for nearby Earth analogs\\I. 15 planet candidates found
  in PFS data\footnote{This paper includes data gathered with the 6.5
    meter Magellan Telescopes located at the Las Campanas Observatory, Chile.}}
\author[0000-0001-6039-0555]{Fabo Feng}
\affiliation{Department of Terrestrial Magnetism, Carnegie Institution of
  Washington, Washington, DC 20015, USA}
\author{Jeffrey D. Crane}
\affiliation{Observatories of the Carnegie Institution for Science, 813 Santa Barbara St., Pasadena, CA 91101}
\author{Sharon Xuesong Wang}
\affiliation{Department of Terrestrial Magnetism, Carnegie Institution of
  Washington, Washington, DC 20015, USA}
\author{Johanna K. Teske}
\affiliation{Department of Terrestrial Magnetism, Carnegie Institution of
  Washington, Washington, DC 20015, USA}
\affiliation{Observatories of the Carnegie Institution for Science,
  813 Santa Barbara St., Pasadena, CA 91101}
\affiliation{Hubble Fellow}
\author{Stephen A. Shectman}
\affiliation{Observatories of the Carnegie Institution for Science, 813 Santa Barbara St., Pasadena, CA 91101}
\author{Mat\'ias R. D\'iaz}
\affiliation{Observatories of the Carnegie Institution for Science, 813 Santa Barbara St., Pasadena, CA 91101}
\affiliation{Departamento de Astronom\'ia, Universidad de Chile, Camino El Observatorio 1515, Las Condes, Santiago, Chile}
\author{Ian B. Thompson}
\affiliation{Observatories of the Carnegie Institution for Science, 813 Santa Barbara St., Pasadena, CA 91101}
\author{Hugh R. A. Jones}
\affiliation{Centre for Astrophysics Research, University of
  Hertfordshire, College Lane, AL10 9AB, Hatfield, UK}
\author{R. Paul Butler}
\affiliation{Department of Terrestrial Magnetism, Carnegie Institute of
  Washington, Washington, DC 20015, USA}

\correspondingauthor{Fabo Feng}
\email{ffeng@carnegiescience.edu}


\begin{abstract}
The radial velocity method plays a major role in the discovery of
nearby exoplanets. To efficiently find planet candidates from the data
obtained in high precision radial velocity surveys, we apply a
signal diagnostic framework to detect radial velocity
signals that are statistically significant, consistent in time,
robust to the choice of noise models, and not correlated with stellar
activity. Based on the application of this approach to the
survey data of the Planet Finder Spectrograph (PFS), we report fifteen
planet candidates located in fourteen stellar systems. We find that
the orbits of the planet candidates around HD 210193, 103949, 8326,
and 71135 are consistent with temperate zones around these stars (where liquid water could exist on the surface). With periods of 7.76 and 15.14\,days respectively, the planet candidates around star HIP 54373 form a 1:2
resonance system. These discoveries demonstate the feasibility of automated detection of exoplanets from large radial velocity surveys,
which may provide a complete sample of nearby Earth analogs. 
\end{abstract}

\section{Introduction}
One of the ultimate goals of exoplanet research is to find nearby
Earth-like planets. The radial velocity (RV) and transit methods have
made the main contributions to exoplanet detections. RV measurements for nearby bright stars can be made relatively efficiently and have enabled planets to be discovered around a significant fraction of them. The transit technique is relatively more sensitive to faint and distant stars. Considering the rare occurrence rate of transit events, the RV method is still of great importance for discovering nearby planets, as evidenced by the detection of Proxima Centauri b \citep{anglada16} although the Transiting Exoplanet Survey Satellite (TESS) is poised to find thousands of nearby transit systems (e.g., \citealt{ricker14}). 

Since the Earth is the only planet known to host life, a conservative
path to the detection of extraterrestial biosignatures is to find an
Earth-like planet around a Sun-like star (also called ``Earth
twin''). To detect such signals, we need to have high precision RV
data which is sensitive to about 0.1\,m/s RV variations \citep{mayor14}. For less massive stars like M dwarfs, the signals corresponding to Earth-sized planets in their temperate zones (where liquid water can exist on the surface; \citealt{kopparapu14}) can be as high as
1\,m/s. To be distinguished from the Earth twins, we call these
planets ``Earth analogs''. Thanks to the recent development
of high precision spectrometers such as ESPRESSO \citep{pepe10} and NEID
\citep{schwab16} as well as advanced noise modeling and activity
mitigating techniques \citep{feng17a,dumusque18}, we are moving towards the detection sensitivity of Earth twins though multiple issues related to stellar activity and instrumental stability (e.g., \citealt{fischer16}) still need to be resolved.

The availability of long-term, high-precision spectroscopic
observations is one of the basic requirements in the discoveries of
Earth analogs such as Proxima Centauri b \citep{anglada16}, GL 667 Cc
\citep{anglada13}, and Luyten b \citep{astudillo-defru17}. 
To this end, the Carnegie Planet Finder Spectrograph (PFS;
\citealt{crane10}) is one of the major instruments dedicated to
performing a high-precision spectroscopic survey of nearby
stars. Since its commissioning in 2010, it has made important contributions to the discovery (Proxima Centauri b at 1.3 pc; \citealt{anglada16}, Barnard's star b at 1.8 pc; \citealt{ribas18}) and characterization (GJ 9827 bcd at 30\,pc; \citealt{teske18}) of nearby planets. 

In this work, we use PFS survey data presented in Section
  \ref{sec:data} and describe how it is processed with a signal
  diagnostic framework in Section \ref{sec:selection} and apply it to
  the PFS survey data, and select signals that are statistically
  significant, consistent in time, uncorrelated with stellar activity
  indicators, and robust to the choice of noise models. We then
  identify planet candidates from these signals and discuss the 15 of
  the most significant signals individually in Section \ref{sec:results}. Finally, we conclude in section \ref{sec:conclusion}.

\section{Data}\label{sec:data}
The PFS measures the Doppler shift of stellar spectral lines through
calibration with the spectrum of iodine (e.g., \citealt{marcy92}). The calibration and the
barycentric correction of the spectrum are implemented through the procedures introduced by
\cite{butler96}. The CaII HK (converted to S-index) and the H$\alpha$
lines are extracted from the spectrum to assess the stellar
activity level. The photon noise is calculated through photon counts
by assuming a Poisson distribution. We consider this photon noise as
an indicator of instrumental and/or stellar noise because it can modulate
the uncertainty of RV measurements and thus change the likelihood of a
periodic signal.

For all PFS targets, we find the relevant astrometry, systematic RV,
and luminosity from {\it Gaia} DR2 \citep{gaiaDR2} through
cross-match using a search cone of 2$'$. The {\it Gaia} source corresponding to a PFS target is identified by selecting the brightest star among all matched
sources. This approach works because most PFS targets are
stars which are bright and with an apparent visual magnitude of
less than 15. The mass of a star is estimated through the
mass-luminosity functions introduced by
\cite{malkov07}, \cite{eker15} and \cite{benedict16}. The stellar type of each star is found by cross-matching the PFS targets with the Simbad database
\citep{wenger00}.

\section{Initial selection of planet candidates}\label{sec:selection}
We define a signal diagnostic framwork to identify signals in the PFS RV data. The following steps are used to search and constrain signals. 
\begin{itemize}
\item We calculate the Bayes factor periodograms (BFPs) for activity
  indices using Agatha \citep{feng17b} and identify activity signals
  at periods of $P_{\rm activity}$. In the calculation of BFP, we
    use the first order moving average model (MA(1);
    \citealt{tuomi12}) to account for time-correlated noise. Compared
  with traditional periodograms such as the Lomb-Scargle
  periodogram \citep{lomb76,scargle82}, the BFP is able to model
    the excess white noise as well as the red noise in a time series. 
  
\item We calculate the BFPs for the RV data using the white, MA(1), and first auto-regressive (AR(1); \citealt{tuomi13}) noise
  models and identify signals for each noise model
  (called ``BFP signals'' at periods of $P_{\rm BFP}$). Although AR(1)
  might lead to false negatives according to \cite{tuomi13}, we use it
  to test whether an RV signal is noise-model dependent.
  
\item We use the adaptive Markov Chain Monte Carlo (MCMC) algorithm
  (called ``DRAM'') developed by \cite{haario06} for model and
  parameter inferences. Specifically, we launch tempered (hot) chains to
  find the global maximum of the posterior and use non-tempered (cold)
  chains to constrain the {\it maximum a posterior} (MAP) signal. Such a combination of
  hot and cold chains allows the parameter space to be well-explored
  without getting stuck in local maxima.  This approach incorporates
  model complexity by considering prior
  distributions. Our algorithm is similar to the hybrid MCMC
  algorithm developed by \cite{gregory11}. Similar to our previous
  works \citep{feng17a}, we adopt a semi-Gaussian prior distribution
  with zero mean and 0.2 standard deviation for eccentricity to
  account for the eccentricity distribution found in radial velocity planets
  \citep{kipping13} and in transit systems \citep{kane12,VanEylen18}. Such a broad semi-Gaussian distribution
  allows solutions with relatively high eccentricity but panelizes
  solutions with extremely high eccentricity. We adopt uniform priors for the logarithmic orbital period and for other orbital
  parameters. 

We constrain each RV signal in the data until any additional
  signals do not increase the likelihood significantly. In other
words, a signal is statistically significant only if the inclusion of
it in the RV model leads to a Bayes factor (BF) of larger than 150 or ln(BF)$>5$
  \citep{kass95,feng16}. The BF is the ratio of marginalized likelihoods
  (or evidences) for two models. We derive the BF from the Bayesian
    information criterion following \cite{kass95}. 
Thus the calculation of BF in this work assumes uniform prior distributions
  of model parameters. This BF criterion is found to be optimal compared with
  other information criteria based on analyses of simulated and
  real RV data sets \citep{feng16}. Therefore we use the BF
    criterion to assess the significance of signals.

The optimal noise model is chosen according to the
  model comparison scheme in Agatha \citep{feng17b}. According to the
  comparison of various noise models for both synthetic and real RV
  data sets \citep{feng16}, the MA models are optimal for avoiding false positives and negatives. Hence we compare the lower and higher order MA models by calculating their BFs. We select the most complex model (with
  highest order) that passes the criterion of ln(BF)$>5$. For example, we calculate the maximum likelihoods for the white noise, MA(1), MA(2),
  and MA(3) models for an RV data set. The logarithmic BFs for MA(1)
  and white noise model and for MA(2) and MA(1) are larger than 5
  while the logarithmic BF for MA(3) and MA(2) is less than 5. The
  optimal noise model would be MA(2) since it is complex enough to
  model the time-correlated noise in the data and is simple enough to
  avoid overfitting. 
\item We define a moving time window and calculate the BFP for each
  step. In other words, we calculate BF$(P_i,W_j)$ (or BF$_{ij}$) for
  the $i^{\rm th}$ period for the $j^{\rm th}$ time window ($W_j$) and form a two dimensional
  power spectrum. Since the nubmer of RVs in different time windows
  are different, we normalize ${\rm BF}_{ij}$ for each time window to
  compare the consistency of signals across time
  windows. Specifically, we scale the maximum BF for time window $W_j$
  to one through ${\bf BF'}_{j}={\bf BF}_{j}/{\rm max}({\bf BF}_{j})$ where ${\bf BF}=\{{\rm BF}_{1j},{\rm BF}_{2j},{\rm BF}_{3j},...\}$. To
optimize the visualization of the significance of signals as a
function of time, the window sizes and steps are chosen according to
the sampling and size of the data as well as the period of target
signal. For example, for a set of 100 RVs sampled uniformly over a
time span of one year, we can divide the time span into 12 bins and
calculate the BFP for the RVs in each bin or time window to
investigate the time-consistency of signals with orbital periods less
than 30 days. For long period signals, broader time windows should be
defined. This example provides a rule of thumb for the choice of
window sizes and steps. We call this two dimensional BFP ``moving periodogram''. It is used to check the time-consistency of signals identified by MCMC (called ``MCMC signals'' at periods of $P_{\rm MCMC}$). We refer the readers to \cite{feng17b} for more details. 
\item We assess the overall quality of an MCMC signal. A genuine Keplerian (due to a planet) signal should be
  \begin{itemize}
  \item robust to the choice of noise models. The difference between the period of the MCMC signal and the corresponding BFP signals are less than 10\% (i.e. $0.9P_{\rm BFP}<P_{\rm MCMC}<1.1P_{\rm BFP}$);
  \item not caused by stellar activity. The difference between the
      period of the MCMC signal and the signals with the two
      highest BFs for each activity index are less than 10\% (i.e. $P_{\rm MCMC}<0.9P_{\rm activity}$ or $P_{\rm MCMC}>1.1P_{\rm activity}$);
  \item statistically significant. The MCMC signal passes the ln(BF)
    threshold of 5;
  \item consistent in time. For an MCMC signal at a period of $P_i$,
    the standard deviation of ${\bf BF}'_{i}=\{{\rm BF}'_{i1},{\rm
      BF}'_{i2},{\rm BF}'_{i3},...\}$ should be less than 0.5. 
    \end{itemize}
  \end{itemize}
  The above criteria are aimed at selecting as many candidate
    signals as possible and at removing obvious false positives. Hence
    the signals identified through these criteria will be further
    studied to investigate their origin. To examine the overlap
    between two signals, we adopt a 20\% period window centered at the
    period of one of the two signals. This use of period ratio or
    percentage rather than a constant period is consistent with our
    use of the uniform distribution as the prior of logarithmic
    period. This period window is narrow for long period signals that
    are typically not well constrained but is broad for short period
    ones that are well constrained by the data. Since most of our data sets have a short time span and cannot be used to identify long
    period signals (e.g., a few years), the period window is broad
    enough to exclude most false short period signals although it
    might also exclude real long period signals. On the other hand, RV signals might be the harmonics of activity signals that are outside of the period
    window and thus cannot be rejected through this
    criterion. However, such a scenario is unlikely because we use two
    signals in each activity index to identify overlaps and the
    harmonics are unlikely to pass all other criteria if it is due to activity. 

Although we use a moving periodogram to check the consistency of signals
in time, it may not always show consistent power even if a signal is
genuine because the power in the BFP for a given time window is also
determined by the time span of the window, the sampling of the data, and
the period of the signal. For example, if the period of a signal is longer
than the time span of the whole data set, there would be no time
window that can cover one period and thus a moving periodgram is not
suitable for a consistency test. In this case, our algorithm would
still count the signal satisfying the time-consistency criterion
although further analyses are needed to investigate the nature of
such signals.

Since many of our PFS data sets contain less than 50 RVs,
  we are cautious about whether a signal can be reliably detected
  and whether the instrument is stable enough for the detection of
  long period signals. We use two criteria to deal with these
  problems. First, we use the moving periodogram to check the
  consistency of signals to diagnose the stability of spectrograph and
  to check the time consistency of signals. Second, we avoid
  overfitting by comparing the null hypothesis and the planet
  hypothesis in the Bayesian framework. We use ln(BF)$>$5
  \citep{kass95,raftery95} to select the best model. Since the null
  hypothesis is typically favored by BF \citep{kass95}, we expect few
  false positives in our automated signal identification. A similar
  conclusion has been drawn by \cite{feng16} based on a comparison of various information criteria in the analyses of simulated and real RV data
  sets. Considering that most short-period Keplerian orbits are not eccentric
  \citep{kipping13}, there are typically less than five efficient
  free parameters for a Keplerian orbit although we calculate the BF using five
  parameters to penalize complex models. Hence the BF criterion is conservative enough to avoid overfitting. 

  The application of the diagnostic procedure to 534 PFS data
sets leads to an identification of 480 periodic signals. We assign the quality
flag of ``A'' to a signal if it satisfies all of the four criteria,
``B'' if it fails to pass one criterion, ``C'' if it fails to pass two
criteria,  ``D'' if it only passes one criterion, and ``E'' if it
satisfies no criterion. By the automated analyses, we find 42, 173,
192, 67, and 5 signals with quality ``A'', ``B'', ``C'', ``D'', ``E'',
respectively. We further investigate the  ``A'' and ``B'' quality
signals and select 15 planetary candidates to report. These targets do
not have available data from other instruments and thus are suitable for
our automated algorithm, which is aimed at identifying signals in RVs measured
by a single instrument. We will design an updated algorithm for
automated analyses of data sets from multiple instruments in the upcoming work.
The physical and observational properties for the stars in these systems are shown in Table \ref{tab:star}. 
\begin{table*}
\caption{Physical parameters and observation parameters for PFS
  targets reported in this paper. The astrometric
 parameters and the effective temperature are provided by {\it Gaia} DR2
\citep{gaiaDR2}. The stellar mass is derived from the {\it Gaia} luminosity using
the mass luminosity relationships in \cite{malkov07}, \cite{eker15} and \cite{benedict16}. The spectral type is
determined through a cross-match of PFS targets with the Simbad database.}
\label{tab:star}
\centering
\begin{tabular}{*9{c}}
\hline\hline
Star&Type&Mass ($M_\odot$)&Temperature (K)&$\alpha$ (deg)&$\delta$
                                                           (deg)
  &$\tilde\omega$(mas)&Time span (day)&Number of RVs\footnote{These RVs
                                        are not binned and multiple RVs may correspond to a single epoch.}\\\hline
HD 210193 & G3V             & $ 1.04 \pm 0.06 $ & $ 5790 _{ -  50 }^{ +  38 }$ & 332.40 & -41.23 & $ 23.67 \pm 0.04 $ & 3161 &26 \\
HD 211970 & K7V           & $ 0.61 \pm 0.04 $ & $ 4127 _{ -  94 }^{ + 149 }$ & 335.57 & -54.56 & $ 76.15 \pm 0.04 $ & 3102 &52 \\
HD 39855 & G8V       & $ 0.87 \pm 0.05 $ & $ 5576 _{ -  46 }^{ +  50 }$ & 88.63 & -19.70 & $ 42.96 \pm 0.03 $ & 2271 &25 \\
HIP 35173 & K2V             & $ 0.79 \pm 0.05 $ & $ 4881 _{ -  81 }^{ +  55 }$ & 109.04 & -3.67 & $ 30.13 \pm 0.06 $ & 3269 &39 \\
HD 102843 & K0V             & $ 0.95 \pm 0.05 $ & $ 5436 _{ -  69 }^{ + 144 }$ & 177.59 & -1.25 & $ 15.91 \pm 0.05 $ & 3009 &36 \\
HD 103949 & K3V             & $ 0.77 \pm 0.04 $ & $ 4792 _{ -  54 }^{ +  66 }$ & 179.55 & -23.92 & $ 37.71 \pm 0.08 $ & 3064 &48 \\
HD 206255 & G5IV/V          & $ 1.42 \pm 0.08 $ & $ 5635 _{ -  99 }^{ +  82 }$ & 325.59 & -50.09 & $ 13.26 \pm 0.03 $ & 3099 &34 \\
HD 21411 & G8V             & $ 0.89 \pm 0.05 $ & $ 5605 _{ - 132 }^{ + 247 }$ & 51.55 & -30.62 & $ 34.30 \pm 0.04 $ & 3217 &31 \\
HD 64114 & G7V             & $ 0.95 \pm 0.05 $ & $ 5676 _{ -  87 }^{ +  32 }$ & 117.98 & -11.03 & $ 31.69 \pm 0.04 $ & 2633 &28 \\
HD 8326 & K2V             & $ 0.80 \pm 0.05 $ & $ 4914 _{ -  32 }^{ +  51 }$ & 20.53 & -26.89 & $ 32.56 \pm 0.05 $ & 3152 &16 \\
HD 164604 & K3.5V         & $ 0.77 \pm 0.04 $ & $ 4684 _{ -  37 }^{ + 135 }$ & 270.78 & -28.56 & $ 25.38 \pm 0.06 $ & 3100 &19 \\
HIP 54373 & K5V             & $ 0.57 \pm 0.03 $ & $ 4021 _{ - 146 }^{ + 226 }$ & 166.86 & -19.29 & $ 53.40 \pm 0.05 $ & 3064 &51 \\
HD 24085 & G0V             & $ 1.22 \pm 0.07 $ & $ 6034 _{ -  53 }^{ +  32 }$ & 56.26 & -70.02 & $ 18.19 \pm 0.02 $ & 3162 &25 \\
HIP 71135 & M1              & $ 0.66 \pm 0.04 $ & $ 4146 _{ - 110 }^{ + 107 }$ & 218.22 & -52.65 & $ 30.90 \pm 0.05 $ & 3009 &44 \\
\hline
\end{tabular}
\end{table*}

\section{Results}\label{sec:results}
The parameters of planet candidates are inferred from the posterior
samples drawn by DRAM chains and are shown in Table
\ref{tab:planet}. The phase curves of all planet candidates and
activity signals are shown in Fig. \ref{fig:phase}. In the calculation
of BFPs for noise models, we do not use Gaussian process (GP) as many
previous studies did \citep{haywood14,rajpaul15} since GP could lead to
false negatives according to recent studies
\citep{feng16,dumusque16a, ribas18}. Moreover, an appropriate kernel
for activity modeling is typically not known if the rotation period and activity
life span is not well determined. We perform uniform analysis of PFS
targets in this work, and could explore incorporating Gaussian
processes informed by photometric data available in the future. Hence
we cannot determine the rotation periods of most stars except in the
cases that rotation-induced activity signals can be found both in activity indices and in RVs. 

\begin{figure*}
  \centering
  \includegraphics[scale=0.8]{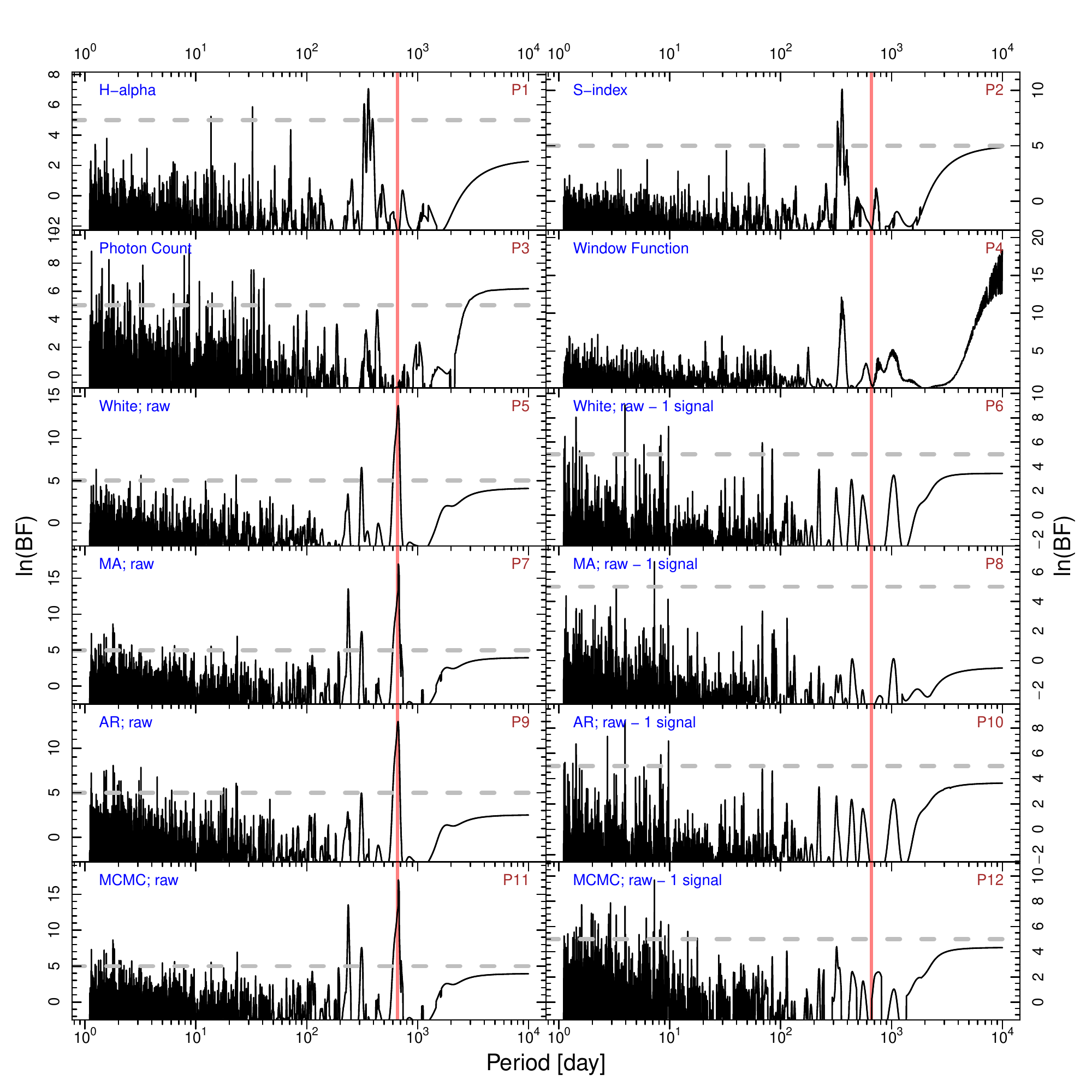}
  \caption{BFPs for HD 210193. The BFPs are calculated for the
    activity indices, and for the white noise model, MA(1), and AR(1) for
    the raw RV data and RVs with sinusoidal signals subtracted. The BFPs for
    activity indices are calculated using the MA(1) model. The window
    function is calculated using the Lomb-Scargle periodogram
    \citep{lomb76,scargle82}. The activity indices, window function,
    noise models and data sets are shown in the top left corners. Each
    panel is denoted by ``Px'' in the top right corner where ``x'' is
    a natural number. The RV signal at a period of 650\,days is marked
    by the red vertical lines. The panels denoted by ``MCMC'' show the
    BFPs for the raw data subtracted by the best-fit Keplerian signals
    constrained by MCMC. The grey dashed line denotes the ln(BF)$=5$
    criterion in each BFP. The elements in this figure are also
    applicable for the subsequent figures. The signals in the residual
  BFPs are either noise-model dependent or not identified as
  significant signals through MCMC sampling. }
  \label{fig:BFP_HD210193}
\end{figure*}

\begin{figure*}
  \centering
  \includegraphics[scale=0.5]{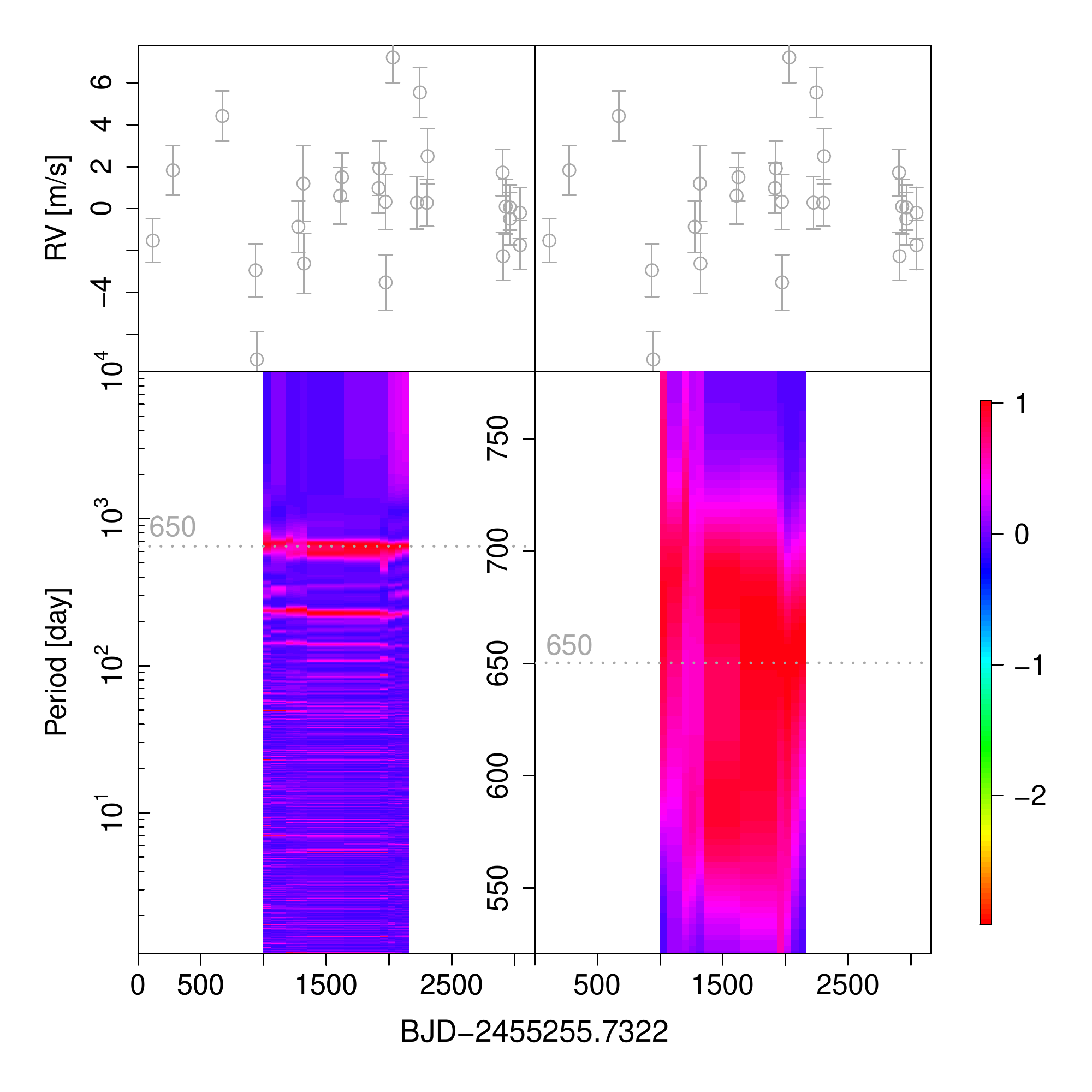}
  \caption{Moving periodogram for the PFS data for HD 210193. The top
    panels show RV data. The bottom right panel is a zoom-in version
    of the bottom left panel which shows the color-coded 2-dimensional
    BFP. {\bf The x axis is the BJD time relative to the first epoch.} The periodogram power is normalized for each moving time
    window so that the BF does not depend on the number of RVs in
    each window. A global color coding is applied to these normalized
    BFs. The time span of each time window is 2036\,days. The window
    move to cover the full data set within 20 steps. In the
    calculation of BFPs, we ignore the eccentricity of signals. This
    is not likely to change the time-consistency of MCMC signals
    although such assumption may alternate signal period and amplitude slightly. }
  \label{fig:MP_HD210193}
\end{figure*}

The RV signal corresponding to a candidate is typically consistently
found in different data chunks, as shown in the moving periodogram
(e.g., see Fig. \ref{fig:MP_HD210193} for HD 210193). As mentioned in section
\ref{sec:selection}, the moving periodogram may not be suitable for all
signals, thus we only show them for the RV signals that need further
confirmation. We also calculate the BFPs for activity indices and do
not find any overlap between activity signals and the Keplerian signals. We display these in a series of plots from
Fig. \ref{fig:BFP_HD210193} onwards where subplots P1 is for H$\alpha$
etc.. The 14 PFS data sets are shown in Table \ref{tab:data}.%

We discuss the results for individual cases as follows.
\begin{table*}
\caption{Parameters for planet candidates. The mass, semi-major axis,
  period, RV semi-amplitude, eccentricity, and mean anomaly at the
  reference epoch are denoted by $M_p$, $a$, $P$, $K$, $e$, $\omega$,
  and $M_0$, respectively. The mean and standard deviation of each
  parameter are estimated from the posterior samples drawn by
  MCMC. For each parameter, the value at the MAP and the uncertainty interval defined by the 1\% and
  99\% quantiles of the posterior distribution are shown below the
  values of mean and standard deviation. The orbit of HD 164604 b is
  significantly eccentric since $e=0$ is more than 3$\sigma$ away from
  the mean. The orbital eccentricities for all other planet candidates are
  consistent with zero.}
\label{tab:planet}
\centering
\begin{tabular}{*8{l}}
\hline\hline
Planet&$M_p$ ($M_\Earth$)& $a$ (au) & $P$ (day) &$K$ (m/s)&$e$&$\omega$ (deg) & $M_0$ (deg) \\\hline
HD 210193 b&$153.1\pm23.3$&$1.487\pm0.031$&$649.918\pm8.599$&$11.40\pm1.66$&$0.24\pm0.09$&$168.84\pm28.69$&$73.93\pm32.63$\\
&$167.39_{-67.60}^{+41.04}$&$1.488_{-0.076}^{+0.071}$&$650.200_{-19.962}^{+20.030}$&$12.55_{-5.14}^{+2.70}$&$0.28_{-0.26}^{+0.16}$&$171.16_{-93.97}^{+66.34}$&$75.84_{-67.25}^{+104.08}$\\
HD 211970 b&$13.0\pm2.5$&$0.143\pm0.003$&$25.201\pm0.025$&$4.02\pm0.74$&$0.15\pm0.10$&$97.83\pm51.73$&$84.80\pm48.66$\\
&$13.28_{-5.91}^{+5.63}$&$0.143_{-0.007}^{+0.006}$&$25.193_{-0.051}^{+0.067}$&$4.04_{-1.76}^{+1.78}$&$0.07_{-0.07}^{+0.35}$&$78.83_{-75.35}^{+125.57}$&$98.86_{-96.58}^{+89.52}$\\
HD 39855 b&$8.5\pm1.5$&$0.041\pm0.001$&$3.2498\pm0.0004$&$4.08\pm0.71$&$0.14\pm0.11$&$102.97\pm79.77$&$154.27\pm77.93$\\
&$8.89_{-3.90}^{+3.19}$&$0.041_{-0.002}^{+0.002}$&$3.2499_{-0.0010}^{+0.0007}$&$4.21_{-1.76}^{+1.88}$&$0.06_{-0.06}^{+0.44}$&$3.97_{-2.44}^{+261.97}$&$258.85_{-254.96}^{+18.02}$\\
HIP 35173 b&$12.7\pm2.7$&$0.217\pm0.004$&$41.516\pm0.077$&$2.80\pm0.59$&$0.16\pm0.11$&$10.73\pm96.40$&$-7.66\pm91.92$\\
&$13.21_{-6.69}^{+5.94}$&$0.217_{-0.010}^{+0.009}$&$41.475_{-0.137}^{+0.219}$&$2.92_{-1.51}^{+1.27}$&$0.23_{-0.22}^{+0.25}$&$31.91_{-207.31}^{+143.42}$&$-0.89_{-174.01}^{+175.62}$\\
HD 102843 b&$113.9\pm14.5$&$4.074\pm0.270$&$3090.942\pm295.049$&$5.24\pm0.61$&$0.11\pm0.07$&$108.69\pm49.38$&$104.42\pm54.78$\\
&$101.23_{-19.99}^{+48.00}$&$3.793_{-0.315}^{+0.950}$&$2770.962_{-312.973}^{+1069.329}$&$4.80_{-0.93}^{+1.96}$&$0.09_{-0.09}^{+0.21}$&$147.46_{-141.03}^{+60.47}$&$42.44_{-37.78}^{+183.19}$\\
HD 103949 b&$11.2\pm2.3$&$0.439\pm0.009$&$120.878\pm0.446$&$1.77\pm0.35$&$0.19\pm0.12$&$127.74\pm68.28$&$234.49\pm67.94$\\
&$11.66_{-5.69}^{+4.95}$&$0.439_{-0.021}^{+0.019}$&$121.040_{-1.199}^{+0.876}$&$1.82_{-0.89}^{+0.78}$&$0.17_{-0.17}^{+0.35}$&$79.95_{-67.49}^{+259.71}$&$287.79_{-269.95}^{+58.68}$\\
HD 206255 b&$34.2\pm7.1$&$0.461\pm0.009$&$96.045\pm0.317$&$3.92\pm0.80$&$0.23\pm0.11$&$86.92\pm44.90$&$124.46\pm46.09$\\
&$37.21_{-19.35}^{+13.97}$&$0.461_{-0.022}^{+0.020}$&$96.027_{-0.707}^{+0.760}$&$4.30_{-2.32}^{+1.49}$&$0.29_{-0.27}^{+0.22}$&$58.41_{-50.95}^{+141.93}$&$149.46_{-133.56}^{+70.96}$\\
HD 21411 b&$65.9\pm25.6$&$0.362\pm0.007$&$84.288\pm0.127$&$11.47\pm4.33$&$0.40\pm0.15$&$332.42\pm17.74$&$285.65\pm22.51$\\
&$75.58_{-68.06}^{+51.87}$&$0.362_{-0.017}^{+0.016}$&$84.232_{-0.239}^{+0.351}$&$13.95_{-8.55}^{+14.71}$&$0.52_{-0.46}^{+0.19}$&$336.90_{-60.98}^{+21.88}$&$275.56_{-35.60}^{+72.82}$\\
HD 64114 b&$17.8\pm3.5$&$0.246\pm0.005$&$45.791\pm0.070$&$3.33\pm0.64$&$0.12\pm0.08$&$215.30\pm85.53$&$209.15\pm85.31$\\
&$17.97_{-8.33}^{+8.01}$&$0.246_{-0.012}^{+0.011}$&$45.799_{-0.172}^{+0.156}$&$3.33_{-1.57}^{+1.52}$&$0.00_{-0.00}^{+0.38}$&$107.07_{-89.00}^{+249.84}$&$342.18_{-323.83}^{+14.14}$\\
HD 8326 b&$66.6\pm19.6$&$0.533\pm0.011$&$158.991\pm1.440$&$9.36\pm2.72$&$0.20\pm0.11$&$-44.79\pm36.46$&$272.91\pm39.43$\\
&$74.20_{-52.97}^{+38.76}$&$0.535_{-0.027}^{+0.023}$&$159.662_{-3.970}^{+2.719}$&$10.35_{-6.71}^{+6.55}$&$0.20_{-0.19}^{+0.31}$&$-35.07_{-111.79}^{+35.00}$&$276.95_{-113.13}^{+75.77}$\\
HD 164604 b&$635.0\pm82.3$&$1.331\pm0.029$&$641.472\pm10.129$&$60.66\pm6.97$&$0.35\pm0.10$&$65.40\pm15.11$&$40.01\pm25.82$\\
&$663.82_{-213.66}^{+168.84}$&$1.323_{-0.061}^{+0.075}$&$635.218_{-16.782}^{+30.166}$&$65.04_{-21.11}^{+12.46}$&$0.42_{-0.33}^{+0.17}$&$65.35_{-39.81}^{+35.80}$&$29.06_{-26.12}^{+115.92}$\\
HIP 54373 b&$8.62\pm1.84$&$0.063\pm0.001$&$7.760\pm0.003$&$4.19\pm0.87$&$0.20\pm0.11$&$72.34\pm44.58$&$90.73\pm43.89$\\
&$8.80_{-4.37}^{+4.18}$&$0.063_{-0.003}^{+0.003}$&$7.760_{-0.007}^{+0.006}$&$4.17_{-1.96}^{+2.25}$&$0.10_{-0.09}^{+0.41}$&$65.31_{-63.27}^{+115.57}$&$103.03_{-96.47}^{+84.33}$\\
HIP 54373 c&$12.44\pm2.11$&$0.099\pm0.002$&$15.144\pm0.008$&$4.84\pm0.79$&$0.20\pm0.12$&$122.37\pm60.04$&$242.71\pm57.33$\\
&$12.87_{-5.25}^{+4.60}$&$0.099_{-0.005}^{+0.004}$&$15.145_{-0.020}^{+0.019}$&$5.00_{-2.04}^{+1.73}$&$0.24_{-0.23}^{+0.26}$&$91.65_{-69.59}^{+236.96}$&$268.06_{-232.15}^{+70.26}$\\
HD 24085 b&$11.8\pm3.1$&$0.034\pm0.001$&$2.0455\pm0.0002$&$5.40\pm1.37$&$0.22\pm0.12$&$8.92\pm54.83$&$156.74\pm60.90$\\
&$14.40_{-9.56}^{+4.66}$&$0.034_{-0.002}^{+0.001}$&$2.0455_{-0.0004}^{+0.0005}$&$6.68_{-4.36}^{+2.40}$&$0.31_{-0.30}^{+0.20}$&$7.86_{-166.98}^{+146.67}$&$149.95_{-132.94}^{+180.97}$\\
HIP 71135 b&$18.8\pm4.1$&$0.335\pm0.007$&$87.190\pm0.381$&$3.71\pm0.79$&$0.21\pm0.13$&$115.60\pm71.72$&$223.03\pm73.08$\\
&$19.51_{-10.04}^{+9.07}$&$0.335_{-0.016}^{+0.015}$&$87.186_{-0.875}^{+0.884}$&$3.79_{-1.94}^{+1.87}$&$0.19_{-0.18}^{+0.38}$&$82.29_{-76.63}^{+250.40}$&$253.41_{-243.15}^{+87.17}$\\
\hline
\end{tabular}
\end{table*}

\begin{figure*}
  \centering
  \includegraphics[scale=0.8]{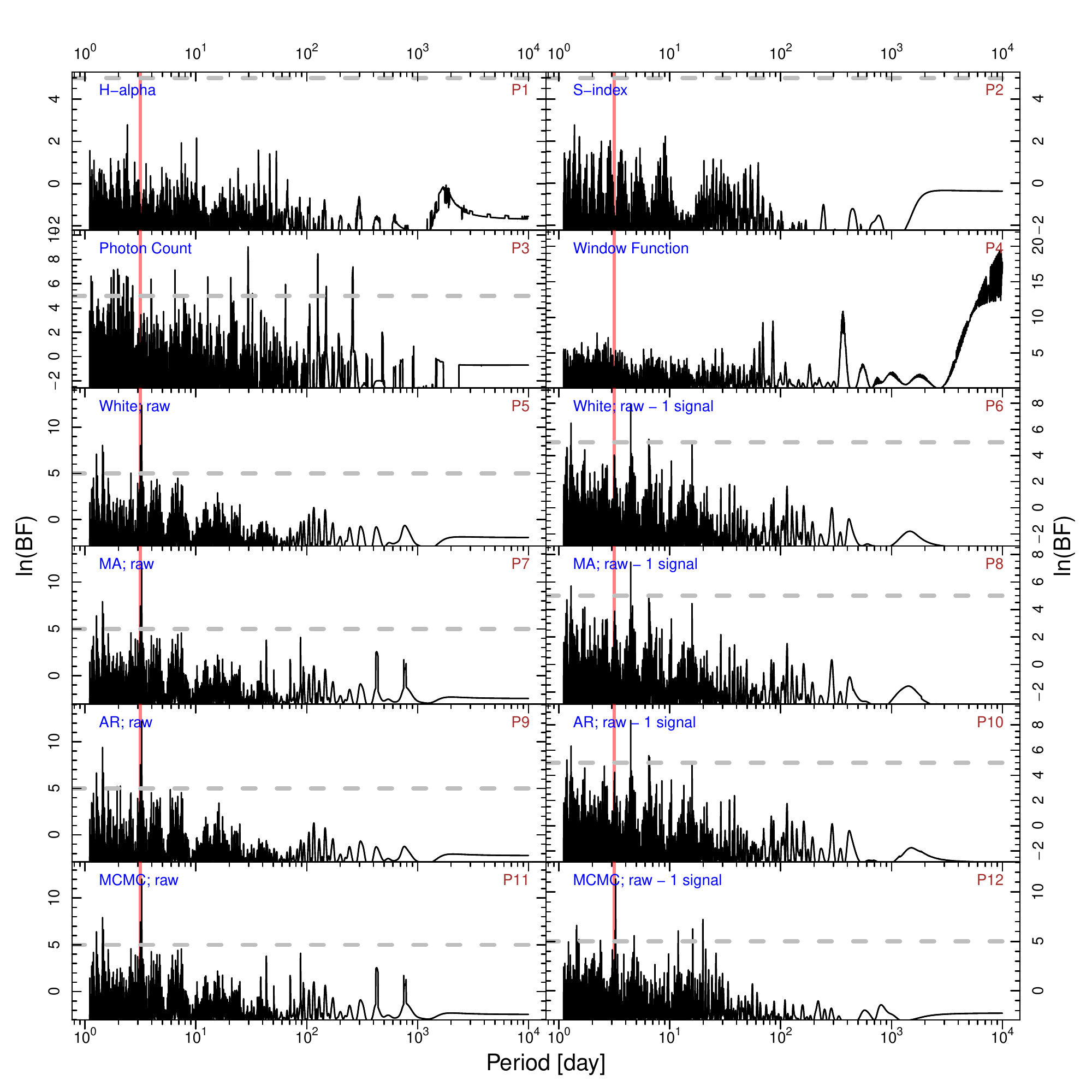}
  \caption{BFPs for HD 39855. The red line shows the signal at a period of 3.25\,days.}
  \label{fig:BFP_HD39855}
\end{figure*}

\begin{figure*}
  \centering
  \includegraphics[scale=0.8]{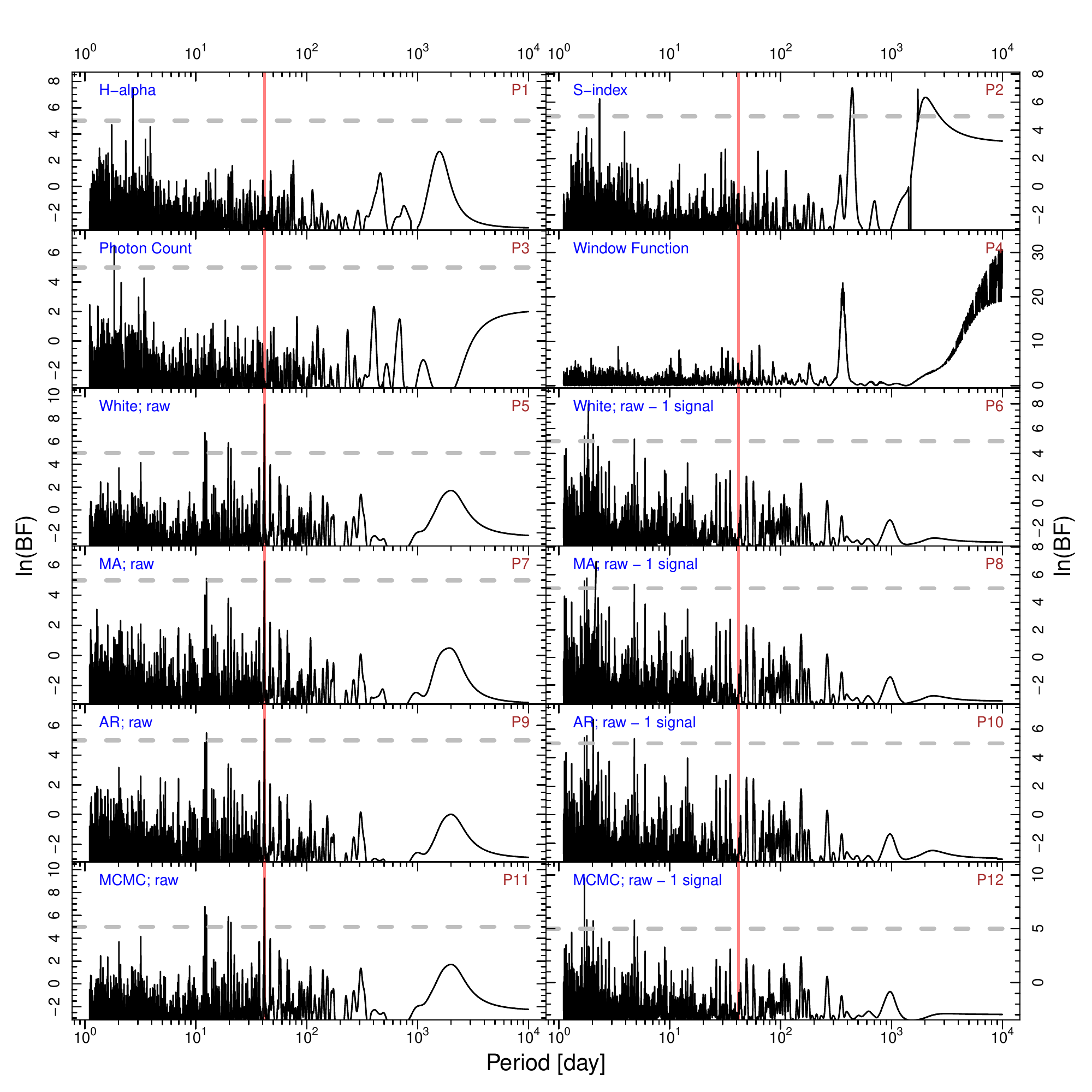}
  \caption{BFPs for HIP 35173. The red line shows the signal at a period of 41.5\,days.}
  \label{fig:BFP_HIP35173}
\end{figure*}

\begin{figure*}
  \centering
  \includegraphics[scale=0.8]{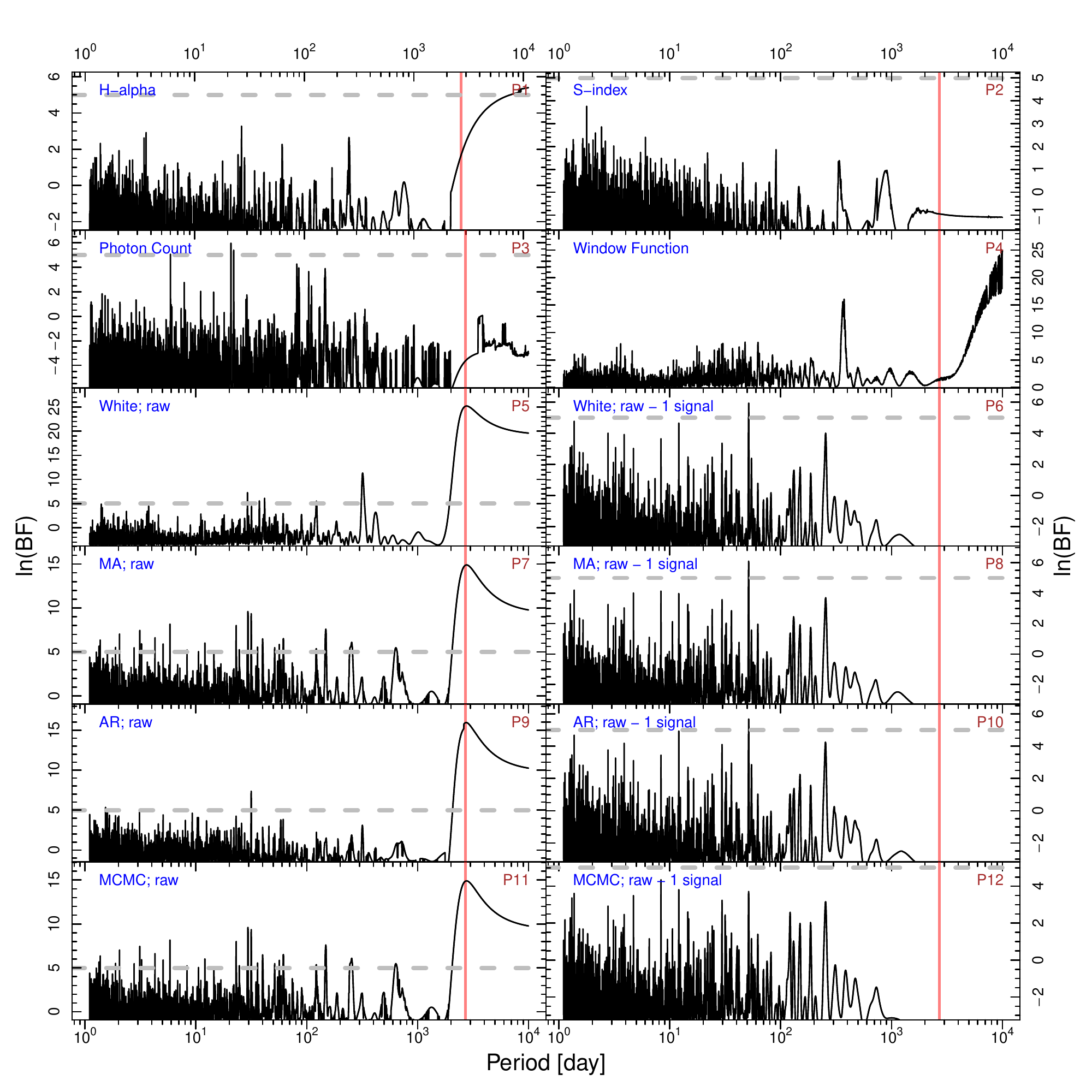}
  \caption{BFPs for HD 102843. The red line shows the signal at a period of about 3000\,days.}
  \label{fig:BFP_HD102843}
\end{figure*}

\begin{figure*}
  \centering
  \includegraphics[scale=0.8]{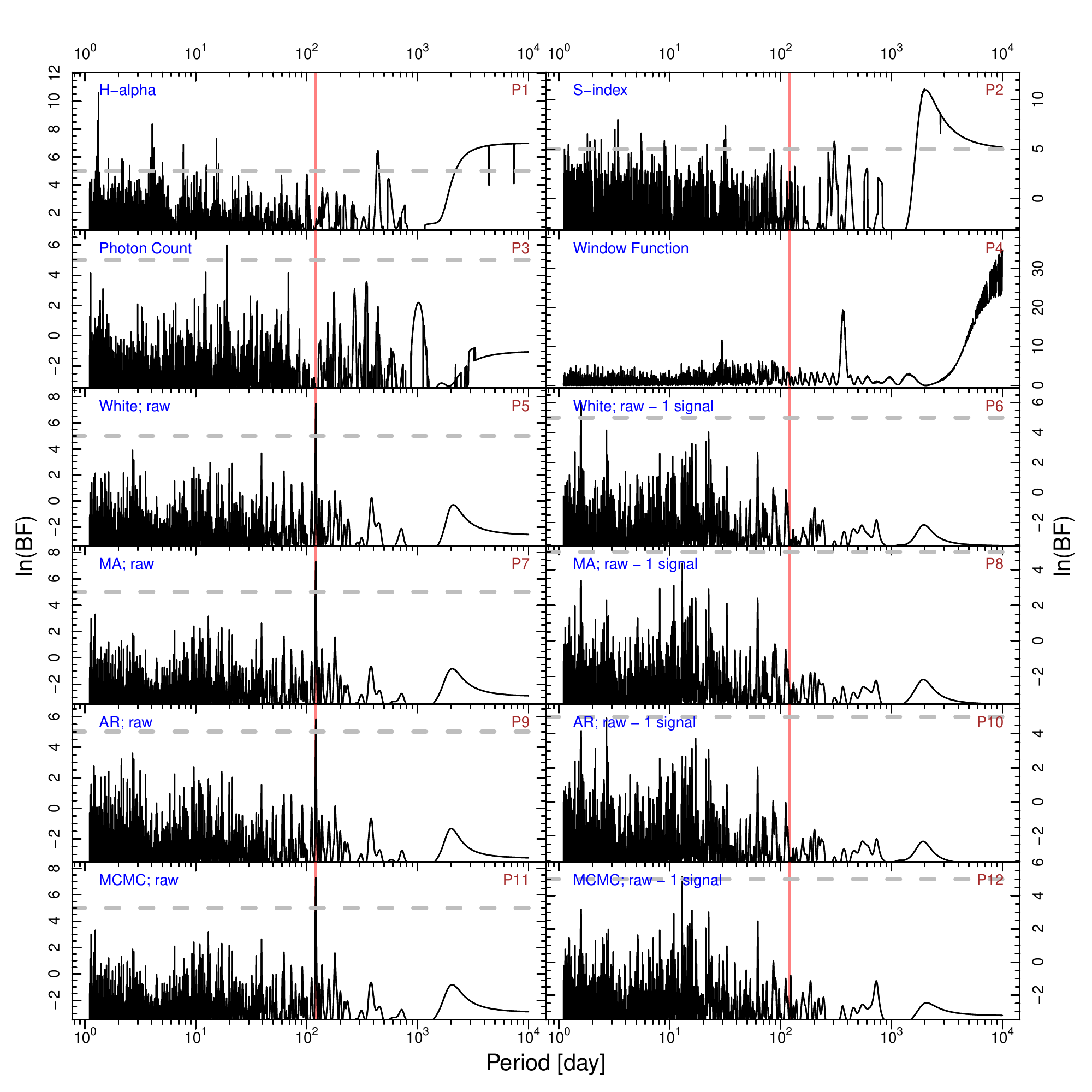}
  \caption{BFPs for HD 103949. The red line shows the signal at a period of 121\,days.}
  \label{fig:BFP_HD103949}
\end{figure*}

\begin{figure*}
  \centering
  \includegraphics[scale=0.8]{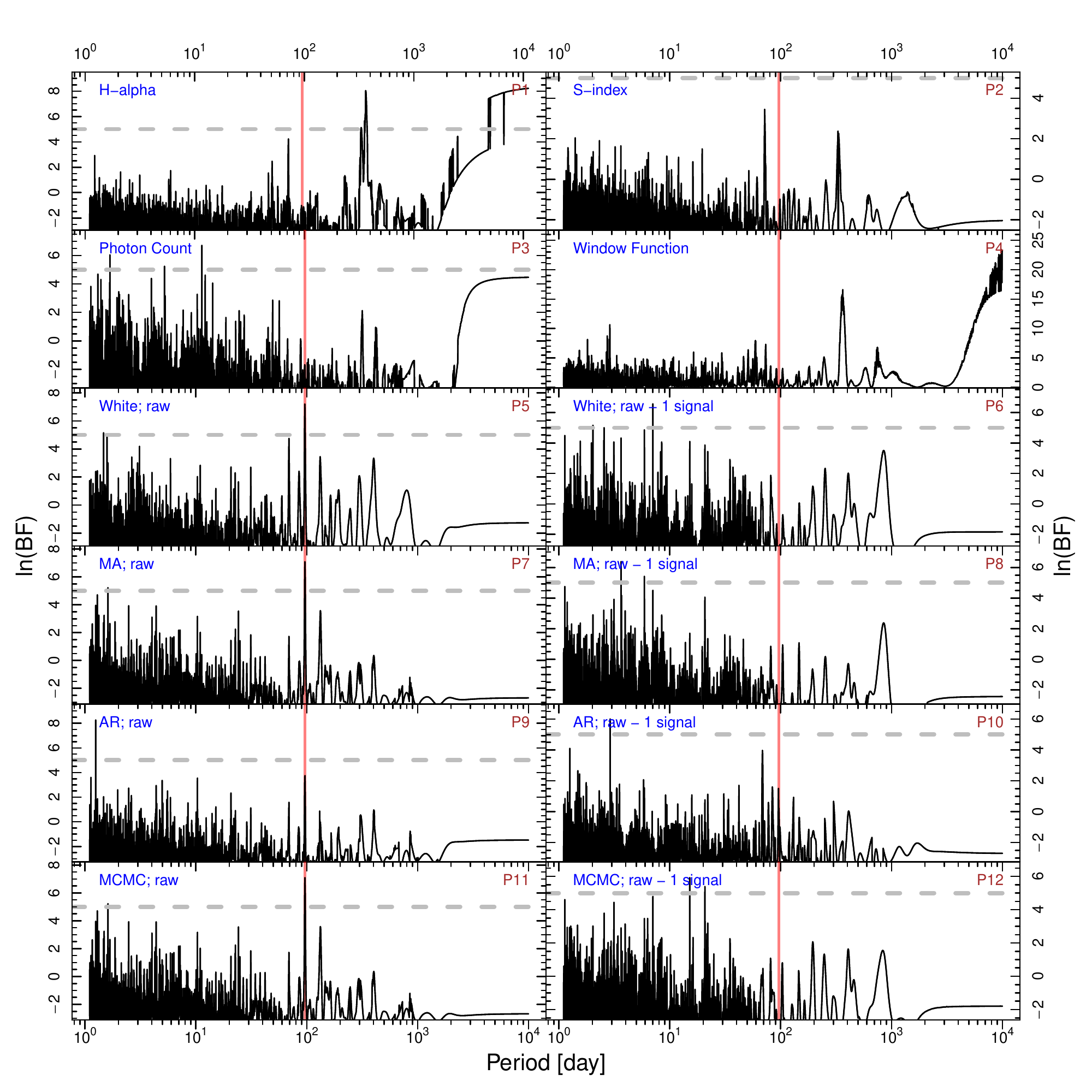}
  \caption{BFPs for HD 206255. The red line shows the signal at a period of 96.0\,days.}
  \label{fig:BFP_HD206255}
\end{figure*}

\begin{figure*}
  \centering
  \includegraphics[scale=0.8]{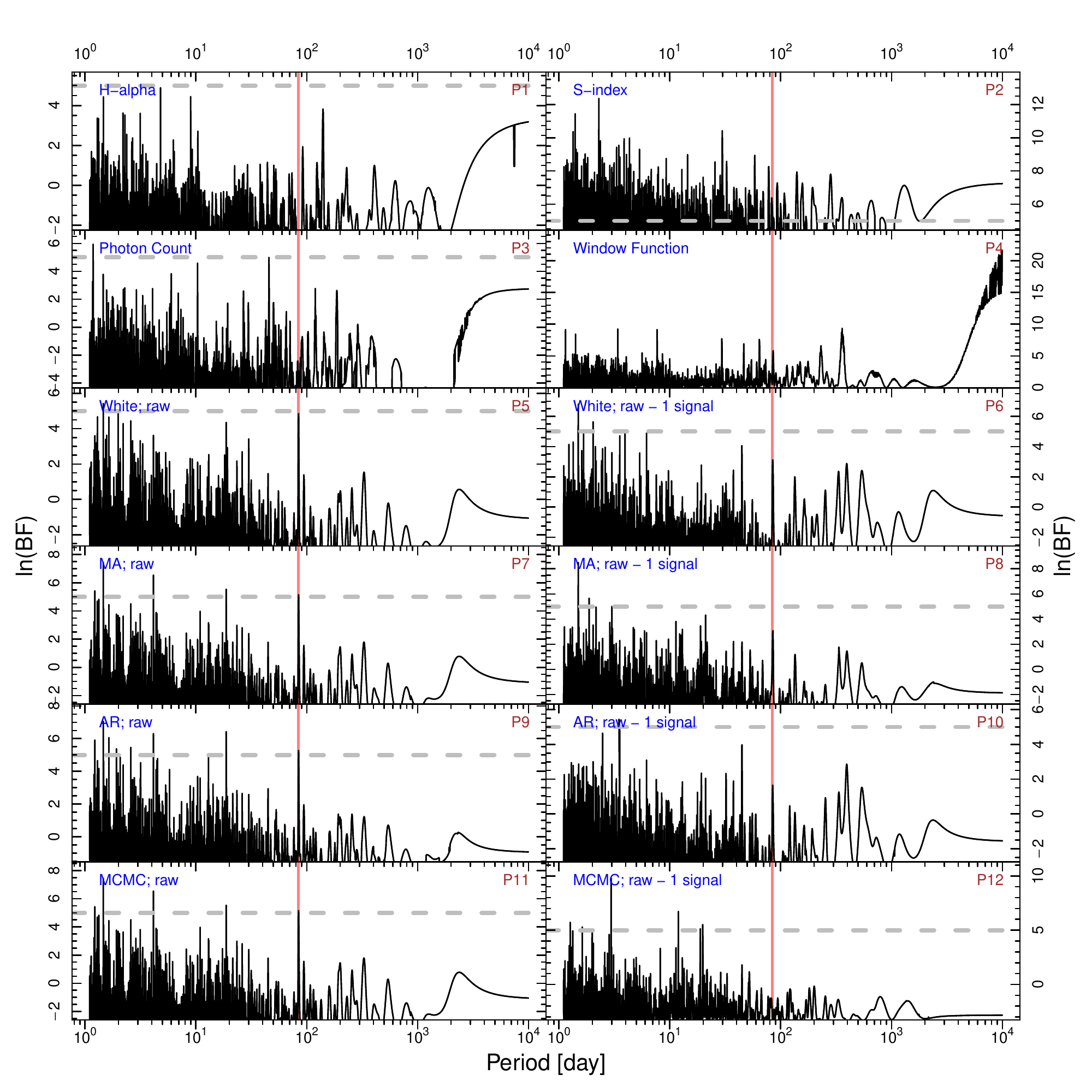}
  \caption{BFPs for HD 21411. The red line shows the signal at a period of 84.3\,days.}
  \label{fig:BFP_HD21411}
\end{figure*}

\begin{figure*}
  \centering
  \includegraphics[scale=0.5]{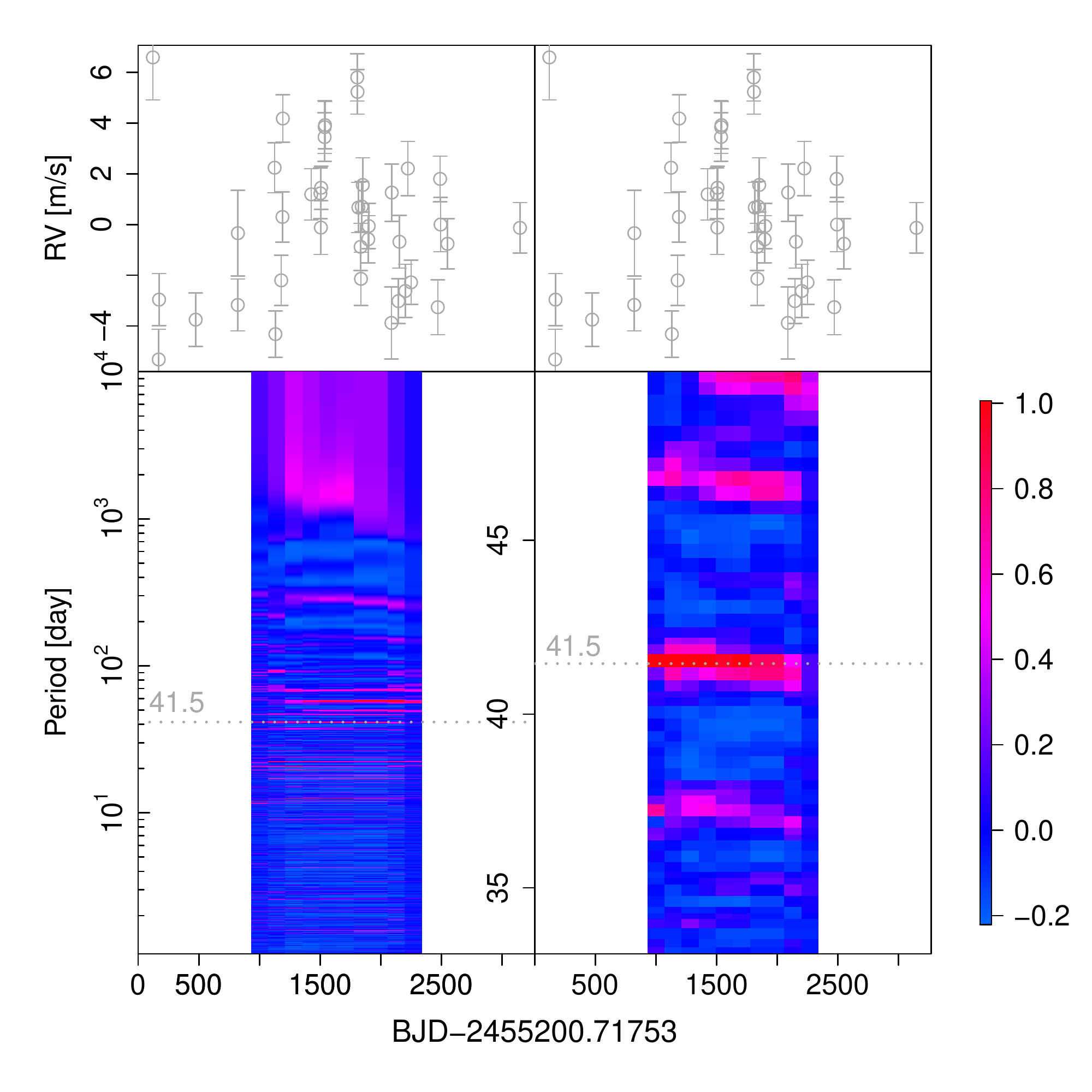}
  \caption{Moving periodogram for HIP 35173 b. The window size is
    2000\,days and the number of steps is 10. The other signals in
    the bottom right panel are the annual aliases of the 41.5\,day signal. }
  \label{fig:MP_HIP35173b}
\end{figure*}

\begin{itemize}
\item \textbf{HD 210193} is a G star with a mass of
  1.04\,$M_\odot$ and distance of 42.25\,pc. The planet candidate has a minimum mass of 153\,$M_\oplus$ and an orbital period of
  about 650\,days. It is a warm giant planet located in the temperate
  zone with the inner and outer boundary corresponding to orbital
  periods of 256 and 926\,days, respectively, according to \cite{kopparapu14}. 

We also calculate the BFPs for activity indices and do not find any
overlap between activity signals and the Keplerian signal, as shown in Fig. \ref{fig:BFP_HD210193}. The phase
curve in Fig. \ref{fig:phase} shows a good fit of
the one-planet model to the RV data in terms of phase coverage and high
statistical significance (i.e. ln(BF)=10.9). The corresponding RV
signal is consistently found in different data chunks, as shown in the moving periodogram (see Fig. \ref{fig:MP_HD210193}). 

There are also other short period signals shown in the BFPs for
residual RVs (P6, P8, P10, and P12 in Fig. \ref{fig:BFP_HD210193}), but they
are either not significant or depend on the choice of noise
models. 
  
\begin{figure*}
  \centering
  \hspace*{-1.1cm}\includegraphics[scale=0.5]{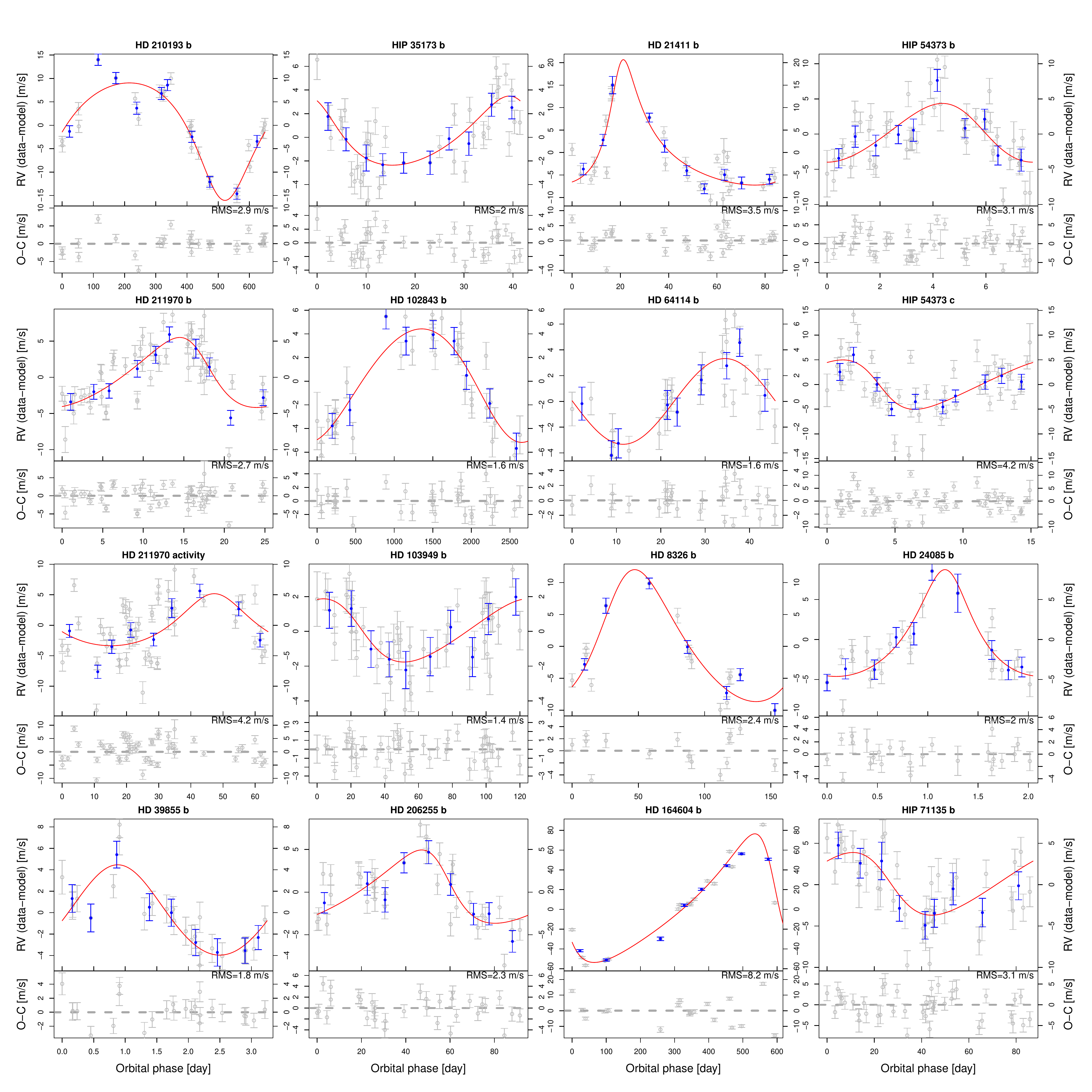}
  \caption{Phase curve and corresponding residuals for all planet
    candidates. The blue error bars show the error-weighted average
    RVs in 10 bins. The best orbital solution is determined by the MAP values of
    orbital parameters. The Root Mean Square (RMS) of the residual RVs
  is shown in each panel. }
  \label{fig:phase}
\end{figure*}

\item \textbf{HD 211970} (GJ 1267) is a K star with a mass of 0.61\,$M_\odot$. It is about 13\,pc away from the Sun and is the nearest star in this reported sample. The planet candidate is consistent with a Neptune with a
  minimum mass of 13.92\,$M_\oplus$ and an orbital period of 25.2\,days. 

We also identify an activity signal at a period of 64.0\,days. This
activity signal shows strong power in the BFPs for RVs and S-index. A
Keplerian fit of the activity signal to the RV data shows non-zero
eccentricity and large residuals with RMS of 4.2\,m/s (see
Fig. \ref{fig:phase}), suggesting an origin due to the quasi-periodic
differential rotation of the star that cannot be properly modeled by
deterministic periodic functions.
Thus the rotation period of this star is probably 64\,days. Since the
true model of activity signal is not known, we report the orbital
parameters for the planet candidate based on the MCMC posterior sampling for one-planet model. 
  
\item \textbf{HD 39855} is a G star with a mass of
  0.87\,$M_\odot$ and a distance of 23.28\,pc. The planet candidate is consistent with a hot super-Earth with a minimum mass of 8.5\,$M_\oplus$ and an orbital period of 3.25\,days. The corresponding RV signal is not evident in the BFPs for  S-index and H$\alpha$ shown in Fig. \ref{fig:BFP_HD39855}. 

\item \textbf{HIP 35173} is a K star with a mass of
  0.79\,$M_\odot$ and a distance of 33.19\,pc. The planet candidate is consistent with a Neptune with a minimum mass of 12.7\,$M_\oplus$ and an orbital period of 41.5\,days. The corresponding RV signal is not evident in the BFPs for the activity indices of S-index and H$\alpha$ shown in Fig. \ref{fig:BFP_HIP35173}. This signal is also consistently strong in various data chunks as seen from the moving periodogram in Fig. \ref{fig:MP_HIP35173b}. 

\item \textbf{HD 102843} is a K star with a mass of
  0.95\,$M_\odot$ and a distance of 62.87\,pc. The planet candidate is
  consistent with a cool Saturn with a minimum mass of 114\,$M_\oplus$ and an orbital period of 3090\,days. Due to its long orbital period, the time span of the data is not long enough to cover
  multiple orbital periods and thus the moving periodogram is not
  appropriate for test of time-consistency. This signal is statistically significant (ln(BF)=7.7 and also see Fig. \ref{fig:phase}) and is unique in the
BFPs for various noise models (see Fig. \ref{fig:BFP_HD102843}). On
the other hand, the BFP for H$\alpha$ shows a signal with a period much longer than the RV signal, probably arising from the magnetic cycle. 

\item \textbf{HD 103949} is a K star with a mass of
  0.77\,$M_\odot$ and a distance of 26.52\,pc. The planet candidate has a minimum mass of 11.2\,$M_\oplus$ and an orbital period of
  121\,days. It is a warm Neptune located in the temperate zone. The corresponding
  RV signal is not evident in the BFPs for the activity indices of
  S-index and H$\alpha$ shown in Fig. \ref{fig:BFP_HD103949}.

\item \textbf{HD 206255} is a G star with a mass of
  1.42\,$M_\odot$ and a distance of 75.40\,pc. The planet candidate is
  consistent with a Neptune with a minimum mass of
  34.2\,$M_\oplus$ and an orbital period of 96.0\,days.

  The corresponding RV signal is unique and has an eccentricity consistent with zero as indicated by the posterior distribution. Thus the extra number of free parameters of the one-planet model with respect to the zero-planet model is better to be set three instead of five in the calculation of BF. The
 former leads to ln$(BF)=8.1$ while the latter to ln(BF)$=4.6$. Although the latter does not pass the ln(BF)$>5$ criterion, we count it as a valid planet candidate due to the above considerations in the calculation of ln(BF) and also because it satisfies all other criteria (see Fig. \ref{fig:BFP_HD206255}). 
  
\item \textbf{HD 21411} is a G star with a mass of
  0.89\,$M_\odot$ and a distance of 29.16\,pc. The planet candidate is consistent with a Neptune with a minimum mass of 65.8\,$M_\oplus$ and an orbital period of 84.3\,days.

  However, the corresponding signal is not as strong as the 18.8\,day
  signal in the BFP (see Fig. \ref{fig:BFP_HD21411}) probably due to its high
  eccentricity ($e=0.4$) and the assumption of circular orbit in the
  calculation of BFP. To confirm the signal, we launch multiple MCMC chains
  and find the same significant 18.8\,day signal, leading to
  ln(BF)$=13.2$. We also constrain the 84.3 and 18.8\,day signals
  simultaneously and find that the two-planet model has a logarithmic
  BF of 2.8 with respect to the one-planet model. Thus we conclude that the
84.3\,days signal is significant while the 18.8\,day signal is not
significant enough to report. The 18.8\,day signal is not see in the BFPs for S-index and H$\alpha$; more follow-up observations are needed to investigate the nature of this signal.

\item \textbf{HD 64114} is a G star with a mass of
  0.95\,$M_\odot$ and a distance of 31.55\,pc. The planet candidate is consistent with a Neptune with a minimum mass of 17.8\,$M_\oplus$ and an orbital period of 45.8\,days. The corresponding RV signal is not evident in the BFPs for the activity indices of S-index in the BFPs for the activity indices of S-index and H$\alpha$ shown in Fig. \ref{fig:BFP_HD64114}.

\item \textbf{HD 8326} is a K star with a mass of 0.8\,$M_\odot$ and
  a distance of 30.71\,pc. The
  candidate is consistent with a planet with a minimum mass of
  66.4\,$M_\oplus$ and an orbital period of 159\,days, consistent with
  the temperate zone around this star. The corresponding RV signal is not evident in the BFPs for the activity indices of S-index and H$\alpha$ shown in Fig. \ref{fig:BFP_HD8326}.

\item \textbf{HD 164604} is a K star with a mass of 0.77\,$M_\odot$ and
  a distance of 39.41\,pc. The candidate is consistent with a Jupiter with a minimum mass of 635\,$M_\oplus$ and an orbital period of 641\,days. The corresponding RV signal is not evident in the BFPs for the activity indices of S-index and H$\alpha$ shown in Fig. \ref{fig:BFP_HD164604}.
  
\item \textbf{HIP 54373} is a K star with a mass of 0.57\,$M_\odot$
  and a distance of 18.73\,pc. There are two planet candidates
  corresponding to a hot super-Earth with a minimum mass of
  8.6\,$M_\oplus$ and an orbital period of 7.76\,days, and a hot
  Neptune with a minimum mass of 12.4\,$M_\oplus$ and an orbital
  period of 15.1\,days. The orbits of these two planets form a 1:2
  resonance. The eccentricities of these two candidate planets are
  consistent with zero since $e=0$ is less than 2 $\sigma$ away from
  the means. The corresponding two RV signals are consistently
  identified in the BFPs for various noise models (see P5, P6, P8, P9,
  P11 and P12 in Fig. \ref{fig:BFP_HIP54373}) and do not show
  significant power excess in the BFPs for S-index and
  H$\alpha$. They are also consistent in time as seen from the moving
  periodograms shown in Fig. \ref{fig:MP_HIP54373b} and
  Fig. \ref{fig:MP_HIP54373c}. The two-planet solution is favored
    over the one-planet solution because ln(BF)$=5.2$. The MAP value
    of the eccentricity for the 15.1\,day signal decreases from 0.30
    for the one-planet model to 0.24 for the two-planet model. Thus the 7.76\,day signal is favored by the data in terms of increasing the goodness
    of fit and reducing the eccentricity of the 15.1\,day signal to a lower and
    thus more reasonable value \citep{kipping13}. According to our analysis of the RV data, there are potentially additional signals, which warrant further observations and analysis. 

\item \textbf{HD 24085} is a G star with a mass of
  1.22\,$M_\odot$ and a distance of 54.99\,pc. The planet candidate is consistent with a hot Neptune with a minimum mass of 11.8\,$M_\oplus$ and an orbital period of 2.04\,days. The corresponding RV signal is not evident in the BFPs for the activity indices of S-index and H$\alpha$ shown in Fig. \ref{fig:BFP_HD24085}.

\item \textbf{HIP 71135} is an M star with a mass of
  0.66\,$M_\odot$ and a distance of 32.36\,pc. The planet candidate is consistent with a Neptune with a minimum mass of 18.8\,$M_\oplus$ and an orbital period of 87.2\,days, consistent with the temperate zone around this star. The corresponding RV signal is not evident in the BFPs for the activity indices of S-index and H$\alpha$ shown in Fig. \ref{fig:BFP_HIP71135}. The signal is consistent in time as seen from the moving periodogram shown in Fig. \ref{fig:MP_HIP71135b}. 
\end{itemize}

\section{Conclusion}\label{sec:conclusion}
We introduce a procedure to diagnose the nature of
  signals in RV data. In this diagnosis framework, we confirm a signal
  as Keplerian if it is statistically significant, consistent in time,
  robust to the choice of noise models, and not correlated with stellar
  activity. We develop an automated algorithm to implement this
  procedure. The application of this algorithm to the PFS data lead to
  an initial identification of about 200 primordial signals with high quality. 

We report 15 planet candidates from these primordial signal
based on analyses of 14 PFS RV data sets that are obtained for
six G stars, seven K stars, and one M star. The masses of planets vary
from 8\,$M_\oplus$ to 153\,$M_\oplus$ and the RV semi-amplitudes vary
from 1.7 to 12\,m/s. The detections of these signals demonstrate the
ability of PFS to discover small planets around nearby stars. 

In particular, we report candidates HD 210193 b, HD 103949 b, HD 8326 b, and HIP
71135 b, which are located in the temperate zones of their stellar
hosts and could potentially host temperate moons. We also report
candidates HIP 54373 b and c, which form a 1:2 resonance. Such a resonance can
stabilize a multiple-planet system for a long period of time, as is
discovered in the TRAPPIST-1 system \citep{gillon16,luger17}.

Since our algorithm only automatically identifies signals in RV data obtained by a single instrument, we choose to report the signals for
targets without other RV data sets available. Thus we do not check the consistency between PFS and other instruments. An updated algorithm will be
developed to automatically analyze multiple RV data sets and to identify potential signals efficiently. Such an algorithm is suitable
for modern RV surveys such as PFS, HARPS \citep{pepe02}, and APF \citep{vogt14}. Our algorithm also provides a diagnostic framework for reliable detections of exoplanets using the RV method. 

\begin{figure*}
  \centering
  \includegraphics[scale=0.8]{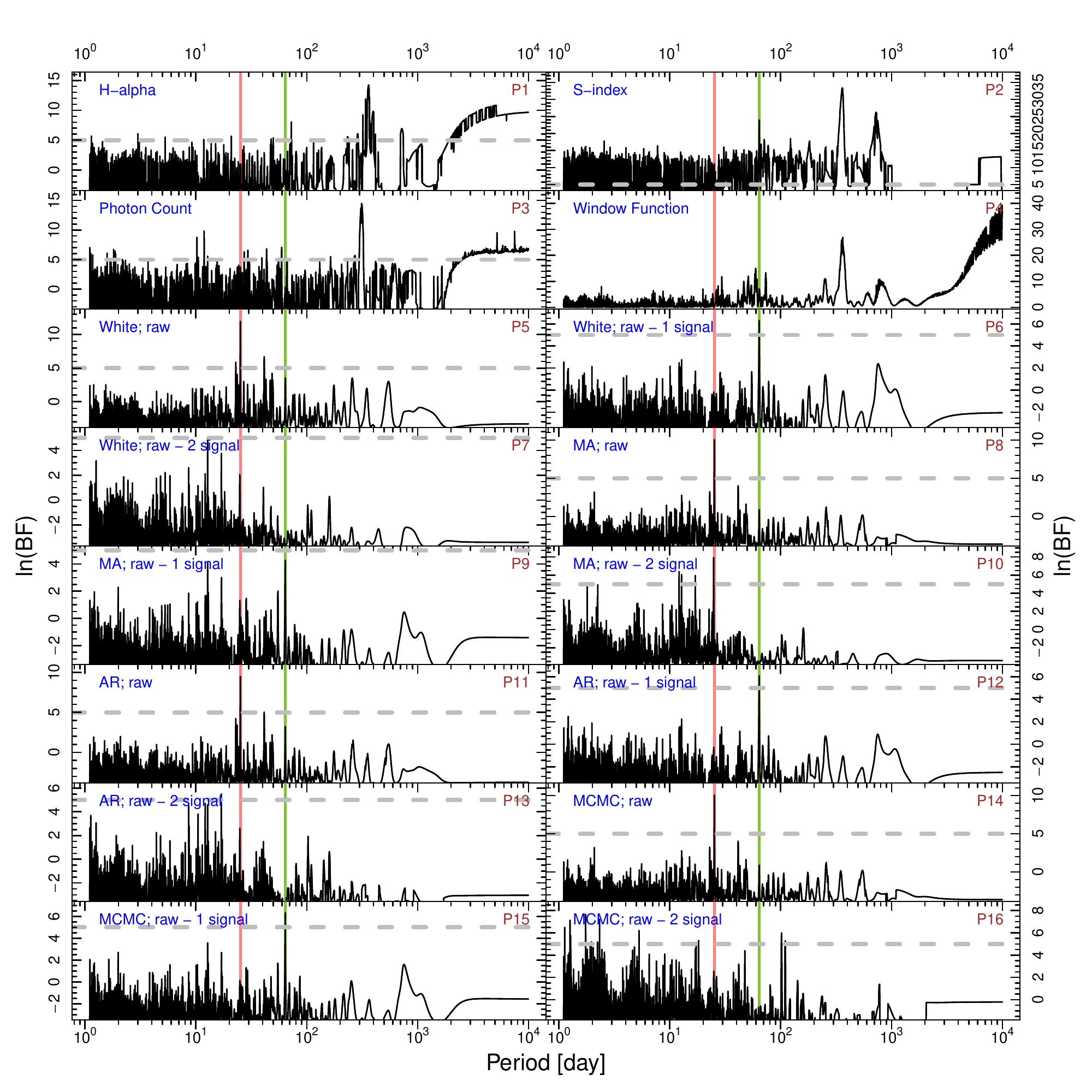}
  \caption{BFPs for HD 211970. The red line shows the Keplerian signal
    at a period of 25.2\,days while the green line shows the 64.0\,day activity signal.}
  \label{fig:BFP_HD211970}
\end{figure*}

\begin{figure*}
  \centering
  \includegraphics[scale=0.8]{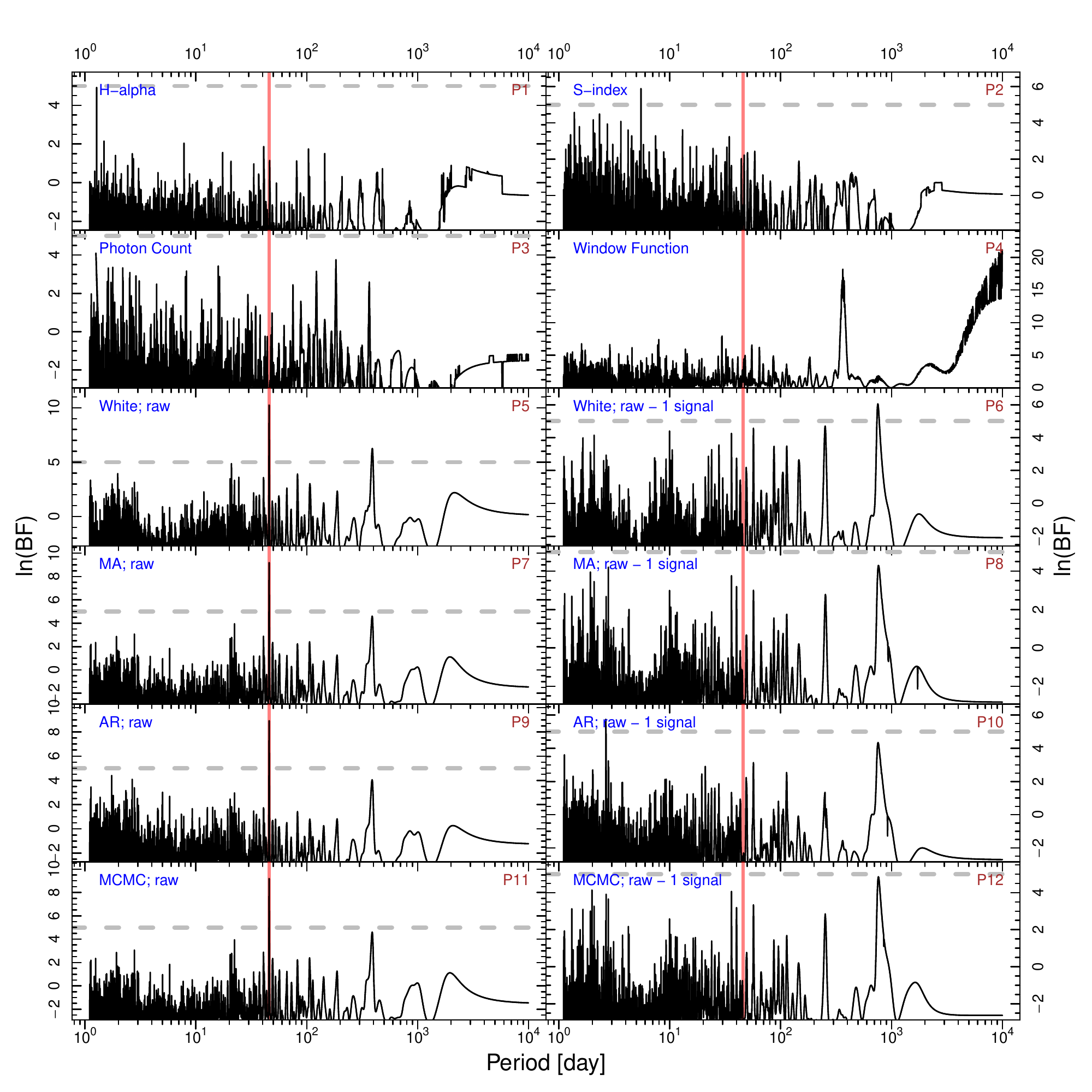}
  \caption{BFPs for HD 64114. The red line shows the signal at a period of 45.8\,days.}
  \label{fig:BFP_HD64114}
\end{figure*}

\begin{figure*}
  \centering
  \includegraphics[scale=0.8]{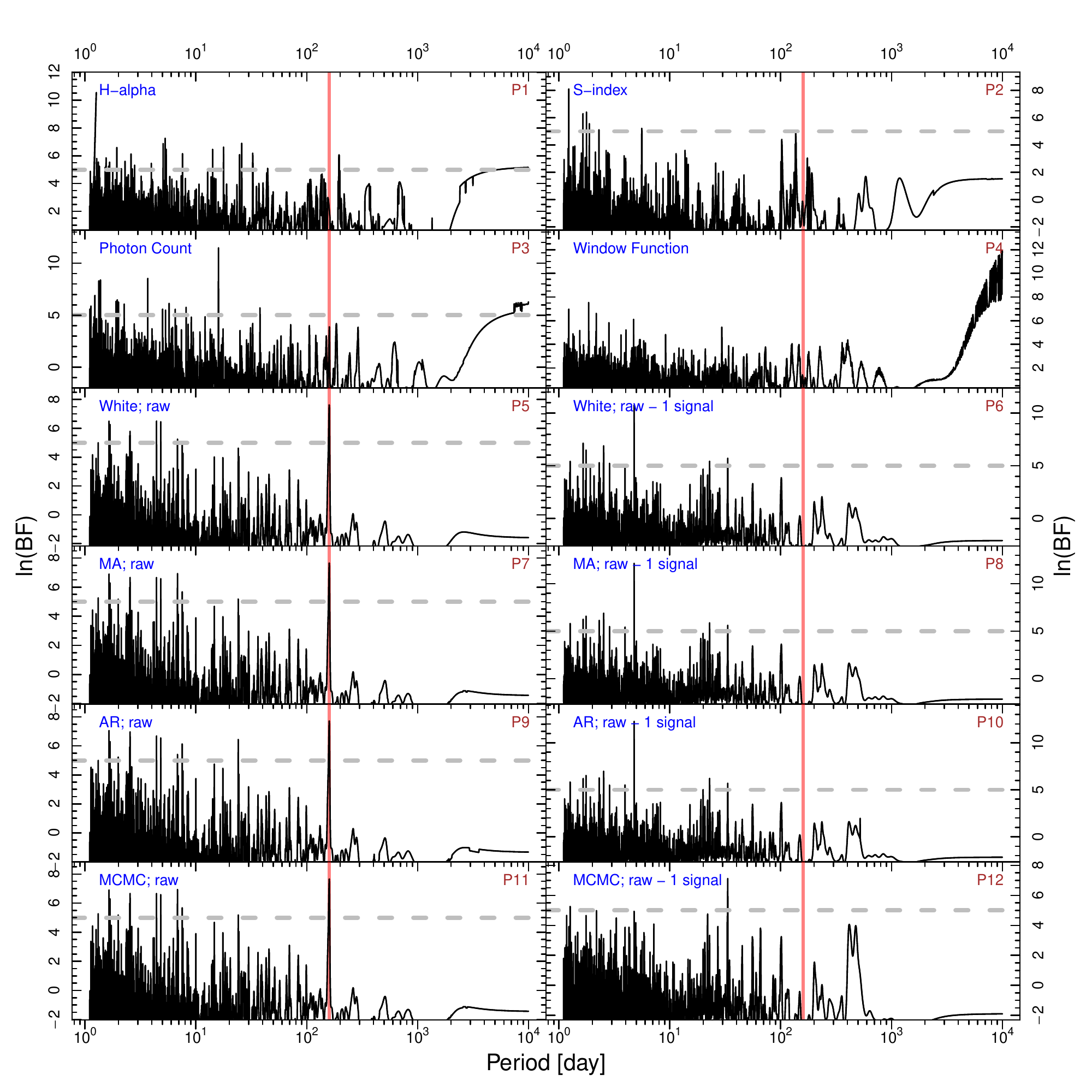}
  \caption{BFPs for HD 8326. The red line shows the signal at a period of 159\,days.}
  \label{fig:BFP_HD8326}
\end{figure*}

\begin{figure*}
  \centering
  \includegraphics[scale=0.8]{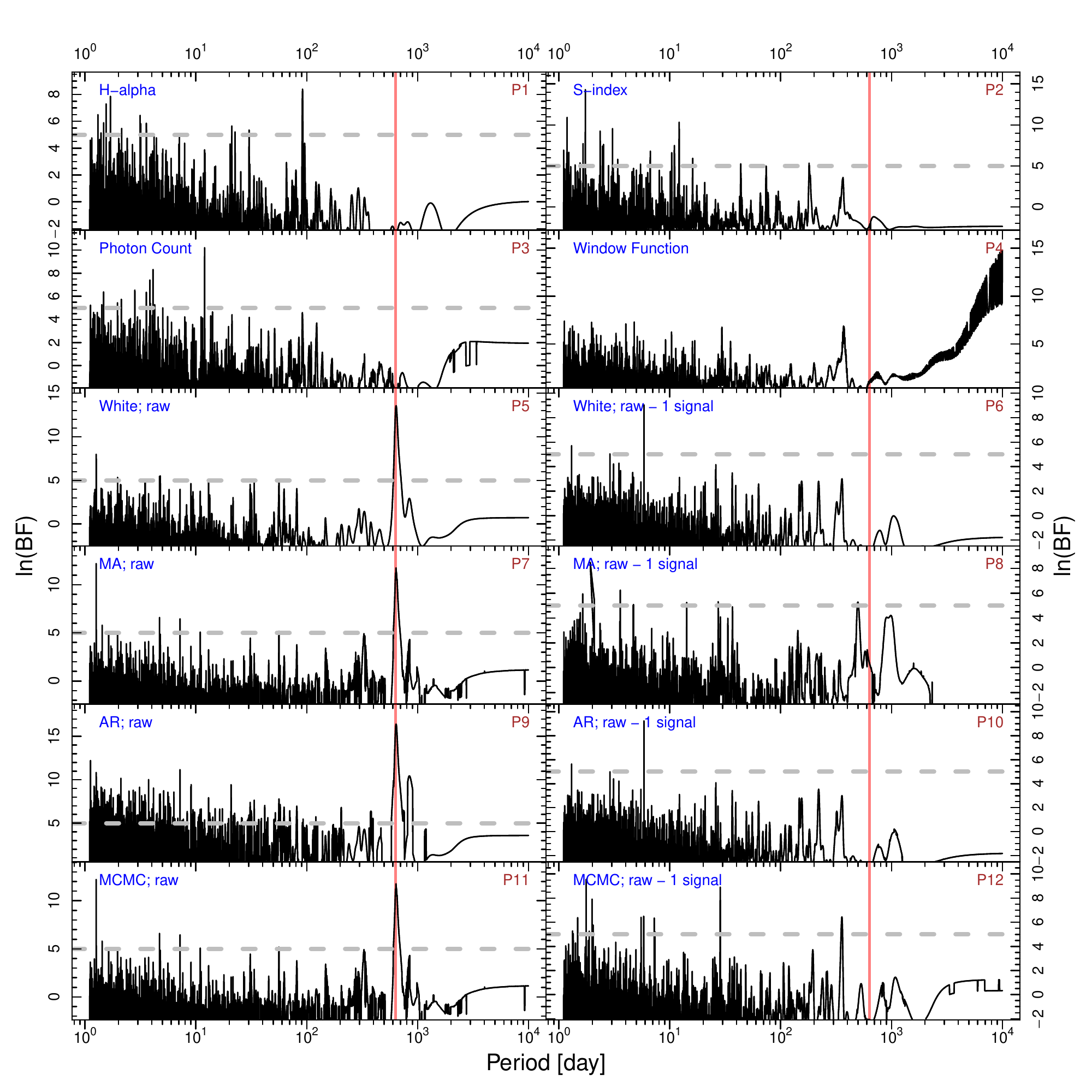}
  \caption{BFPs for HD 164604. The red line shows the signal at a period of 635\,days.}
  \label{fig:BFP_HD164604}
\end{figure*}

\begin{figure*}
  \centering
  \includegraphics[scale=0.8]{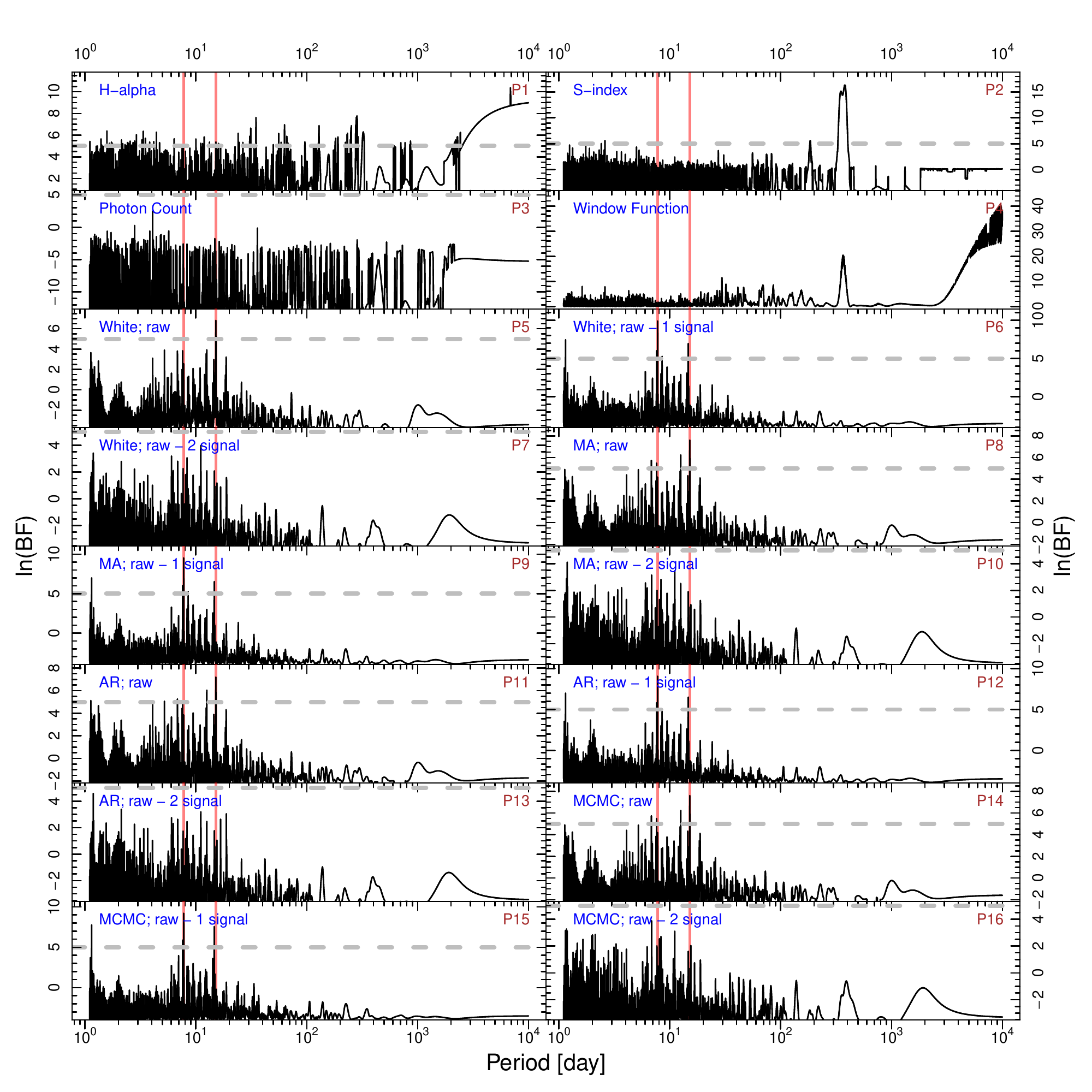}
  \caption{BFPs for HIP 54373. The red lines show the signals at periods of 15.1 and 7.76\,days.}
  \label{fig:BFP_HIP54373}
\end{figure*}

\begin{figure*}
  \centering
  \includegraphics[scale=0.8]{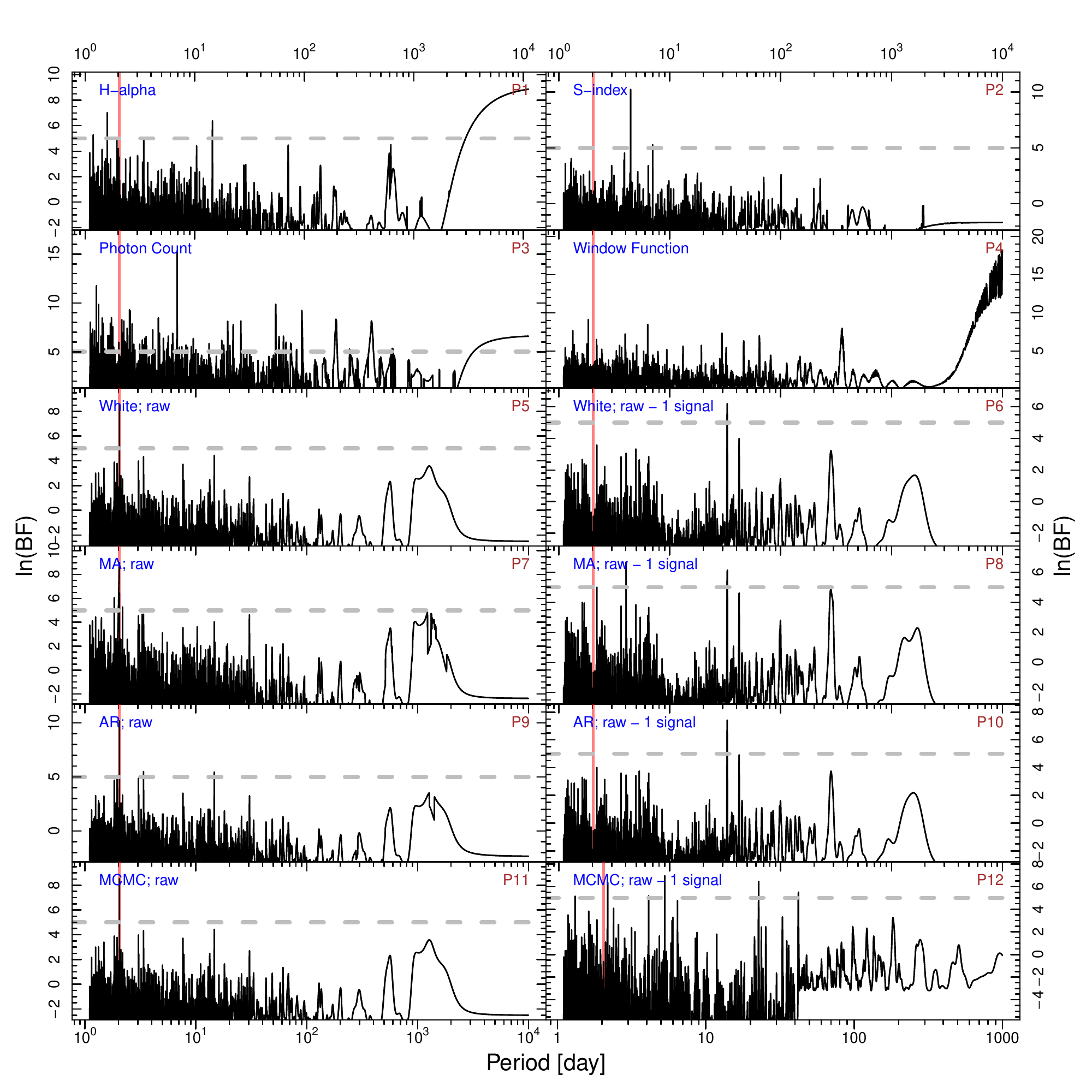}
  \caption{BFPs for HD 24085. The red line shows the signal at a period of 2.05\,days.}
  \label{fig:BFP_HD24085}
\end{figure*}

\begin{figure*}
  \centering
  \includegraphics[scale=0.8]{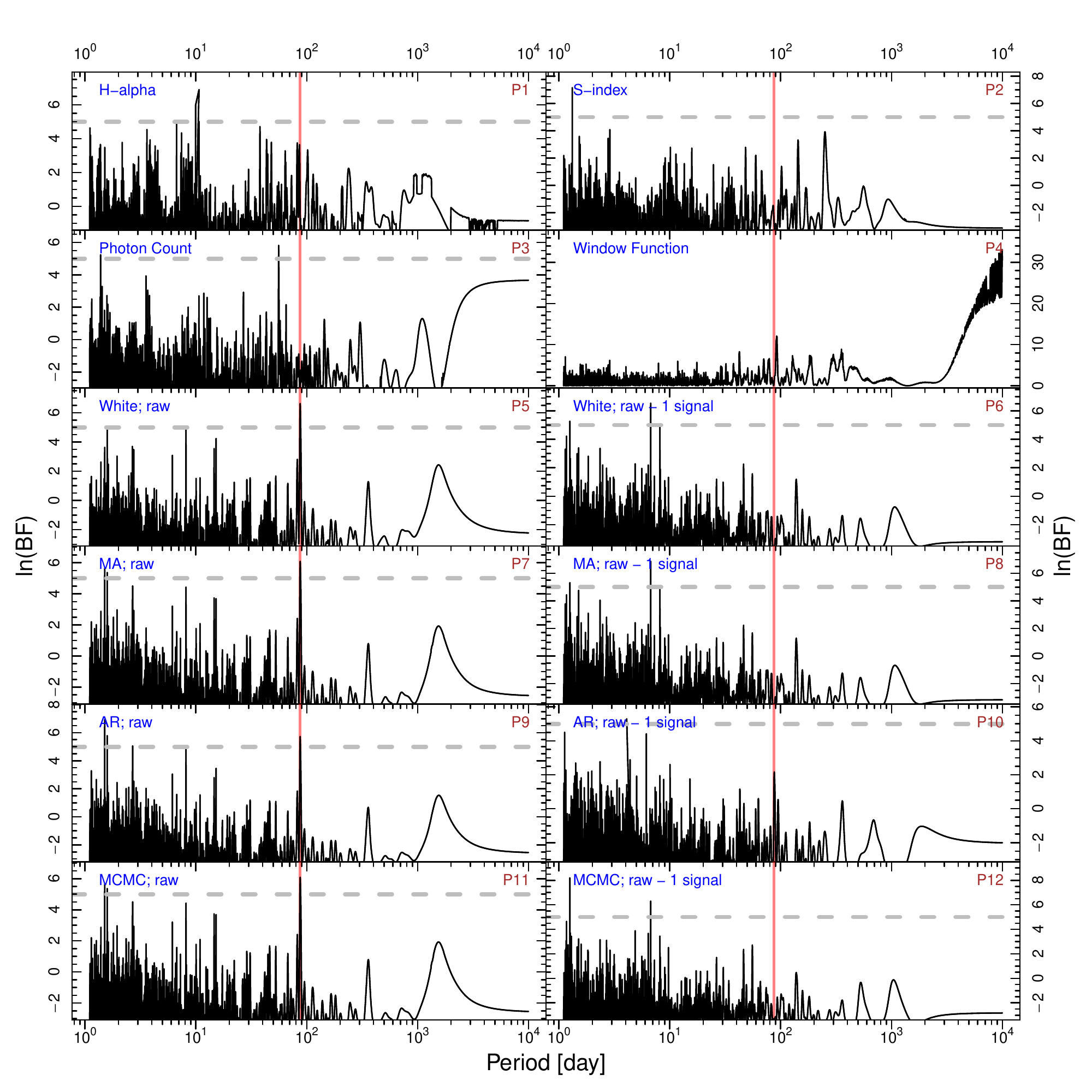}
  \caption{BFPs for HIP 71135. The red line shows the signal at a period of 87.2\,days.}
  \label{fig:BFP_HIP71135}
\end{figure*}

\begin{figure*}
  \centering
  \includegraphics[scale=0.5]{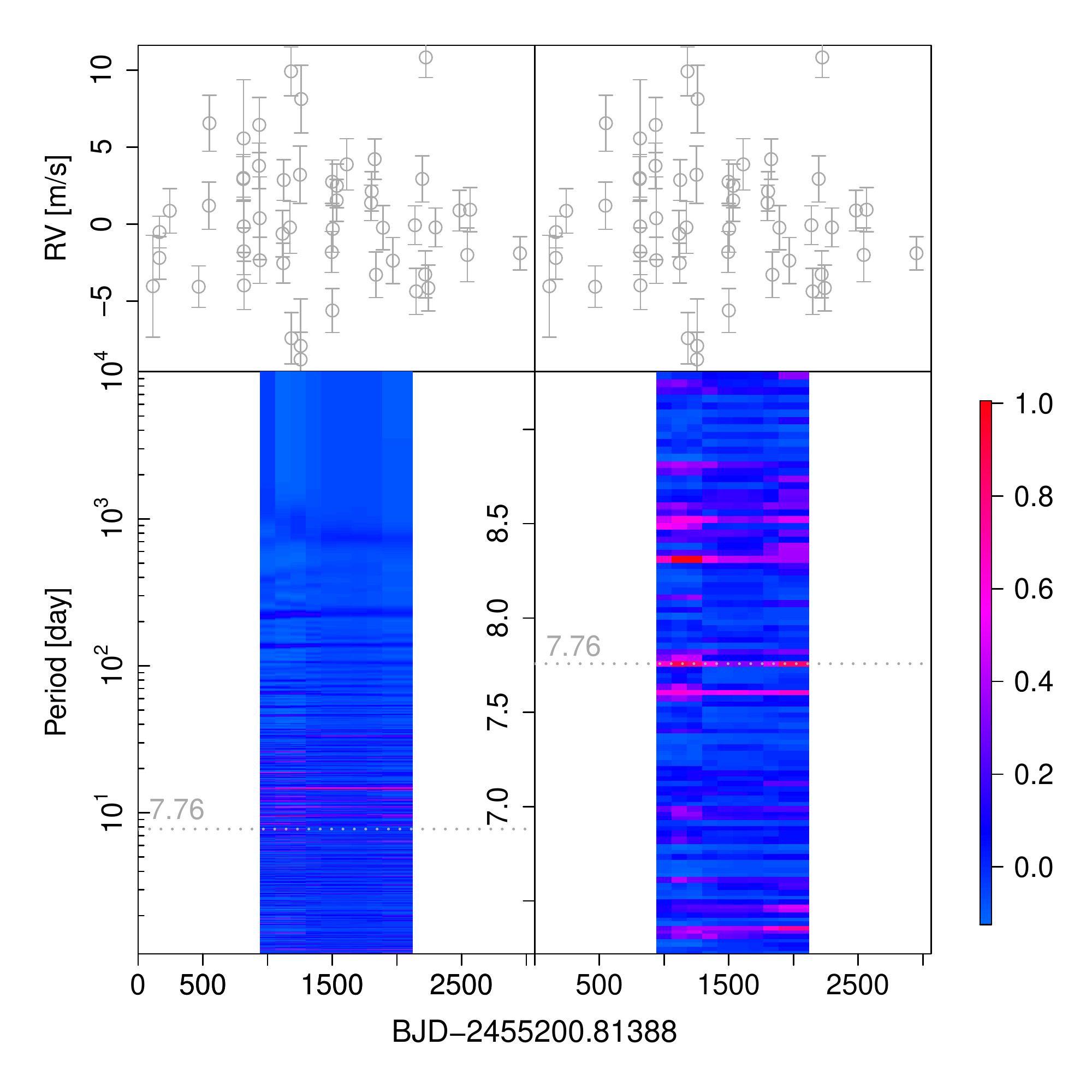}
  \caption{Moving periodogram for HIP 54373 b. The window size is
    2000\,days and the number of steps is 10. The signal near the
    7.76\,day signal is its annual alias. }
  \label{fig:MP_HIP54373b}
\end{figure*}

\begin{figure*}
  \centering
  \includegraphics[scale=0.5]{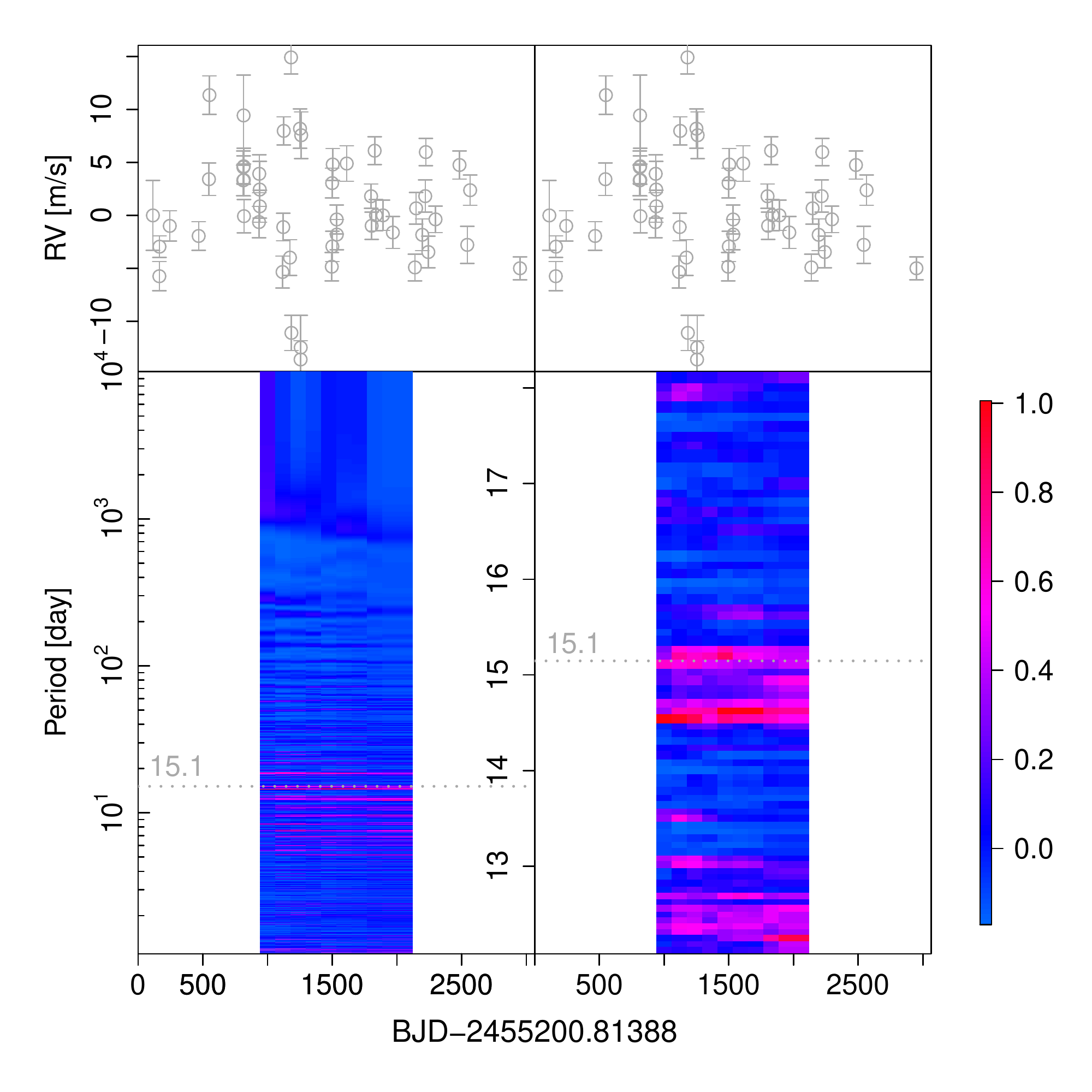}
  \caption{Moving periodogram for HIP 54373 c. The window size is
    2000\,days and the number of steps is 10. The signal near the
    15.1\,day signal is its annual alias. }
  \label{fig:MP_HIP54373c}
\end{figure*}

\begin{figure*}
  \centering
  \includegraphics[scale=0.5]{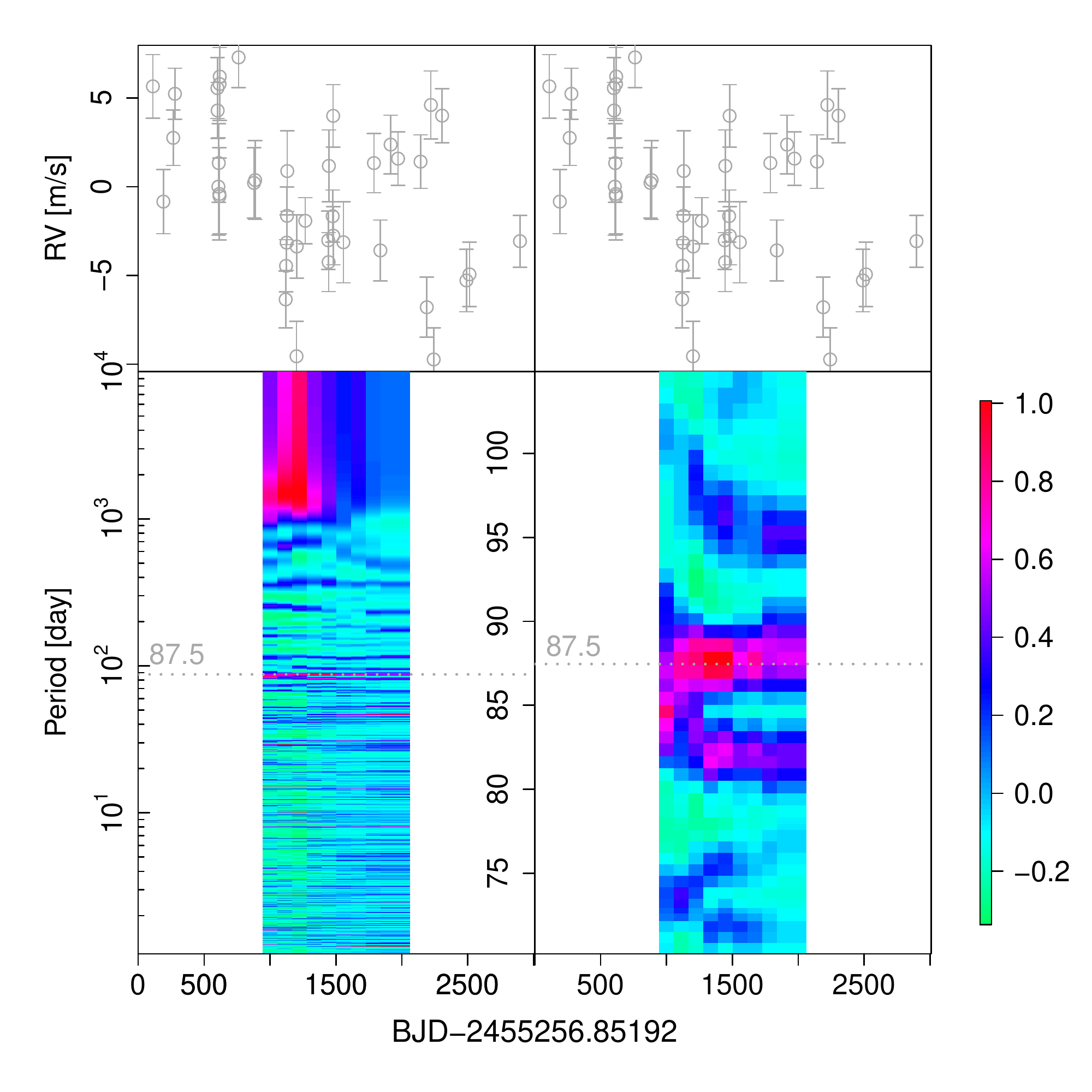}
  \caption{Moving periodogram for HIP 71135 b. The window size is
    2000\,days and the number of steps is 10. The annual alias of
    87.5\,day also show excess power in the bottom right panel. }
  \label{fig:MP_HIP71135b}
\end{figure*}

\clearpage
\startlongtable
\begin{deluxetable*}{ccccccc}
\tablecaption{PFS data for the 14 targets. The nonvalid values in the tables are denoted by -1.\label{tab:data}}
\tablehead{
\colhead{Star}&\colhead{BJD[TDB]} & \colhead{RV} & \colhead{RV error} & \colhead{S-index} & \colhead{H$\alpha$} & \colhead{Photon Count}\\
\colhead{}&\colhead{(day)}&\colhead{(m/s)}&\colhead{(m/s)}&\colhead{}&\colhead{}&\colhead{}}
\startdata
HD 210193&2455255.7322&-2.96&1.03&0.2681&0.03949&17240\\\hline
HD 210193&2455427.74325&10.5&1.19&0.1453&0.0298&20808\\\hline
HD 210193&2455853.59631&-3.87&1.2&0.1457&0.03015&29319\\\hline
HD 210193&2456139.71964&6.04&1.26&0.1408&0.03059&34033\\\hline
HD 210193&2456150.73734&1.68&1.34&0.1459&0.03024&17527\\\hline
HD 210193&2456507.78955&-8.57&1.22&0.1407&0.03049&33751\\\hline
HD 210193&2456551.68262&-0.75&1.8&0.1522&-1&23658\\\hline
HD 210193&2456556.69412&-3.99&1.44&0.1589&0.031&16623\\\hline
HD 210193&2456867.80008&7.54&1.35&0.153&0.02991&29091\\\hline
HD 210193&2456883.72837&7.57&1.15&0.1475&0.03002&41296\\\hline
HD 210193&2457198.88933&-1.32&1.2&0.1479&0.02977&33208\\\hline
HD 210193&2457205.83287&0.43&1.3&0.1532&0.02989&40291\\\hline
HD 210193&2457258.62318&3.76&1.32&0.1588&0.03054&27175\\\hline
HD 210193&2457259.74463&0&1.32&0.1558&0.03054&21843\\\hline
HD 210193&2457321.59187&14.24&1.2&0.1562&0.02954&24402\\\hline
HD 210193&2457529.88662&6.59&1.25&0.1457&0.0302&23762\\\hline
HD 210193&2457554.90868&10.2&1.21&0.1586&0.03064&21573\\\hline
HD 210193&2457614.76154&-1.31&1.13&0.1468&0.03049&36335\\\hline
HD 210193&2457620.6494&0.09&1.32&0.1514&-1&20191\\\hline
HD 210193&2458266.85745&-0.13&1.11&-1&-1&16288\\\hline
HD 210193&2458270.89806&-4.68&1.14&-1&-1&22104\\\hline
HD 210193&2458293.82287&-5.82&1.3&-1&-1&18573\\\hline
HD 210193&2458329.77738&-11.74&1.08&-1&-1&7636\\\hline
HD 210193&2458329.78116&-12.28&1.24&-1&-1&7132\\\hline
HD 210193&2458416.55692&-15.24&1.17&-1&-1&7138\\\hline
HD 210193&2458416.5608&-13.7&1.22&-1&-1&7617\\\hline
\hline
HD 211970&2455255.7322&-2.96&1.03&0.2681&0.03949&17240\\\hline
HD 211970&2455427.75376&4.39&1.01&1.0036&0.05437&24542\\\hline
HD 211970&2455439.7522&4.22&1.16&0.9017&0.05306&20281\\\hline
HD 211970&2455785.66778&-6.38&1.17&0.6735&0.05198&22935\\\hline
HD 211970&2455787.78384&-5.22&1.11&0.7317&0.05236&25121\\\hline
HD 211970&2455790.64196&-0.3&1.15&0.7294&0.05265&20023\\\hline
HD 211970&2455793.66157&-5.06&1.2&0.8489&0.05269&17699\\\hline
HD 211970&2455795.73434&2&1.11&0.724&0.05198&21689\\\hline
HD 211970&2455801.6949&2.91&1.24&0.8221&0.05361&12861\\\hline
HD 211970&2455802.66507&3.1&1.26&0.7252&0.05326&14089\\\hline
HD 211970&2455843.73573&0.23&1.26&0.4834&0.05189&15628\\\hline
HD 211970&2455844.62917&-1.09&1.05&0.6433&0.05282&17331\\\hline
HD 211970&2455845.65818&-3.1&1.07&0.6588&0.05241&16444\\\hline
HD 211970&2455846.6955&-1.07&1.1&0.6576&0.05249&21090\\\hline
HD 211970&2455850.66816&3.63&1.2&0.6687&0.0551&13526\\\hline
HD 211970&2455851.6625&1.67&1.27&0.5961&0.05486&14869\\\hline
HD 211970&2455852.63469&-0.86&1.29&0.6993&0.05403&13643\\\hline
HD 211970&2455853.60615&-1.63&1.43&0.5476&0.05468&10524\\\hline
HD 211970&2456085.91961&-4.78&1.14&0.7777&0.05213&20813\\\hline
HD 211970&2456086.82561&-3.88&1.02&0.8023&0.05236&25862\\\hline
HD 211970&2456087.91929&-5.73&1.45&0.9308&0.05402&10622\\\hline
HD 211970&2456092.87766&0.67&1.07&0.874&0.0524&23615\\\hline
HD 211970&2456141.7073&0.99&1.12&0.8227&0.05365&23515\\\hline
HD 211970&2456147.73551&5.39&1.57&0.8008&-1&9014\\\hline
HD 211970&2456501.7777&6.74&1.24&0.8439&0.05292&18659\\\hline
HD 211970&2456504.84093&1.5&1.86&0.9769&0.05405&6983\\\hline
HD 211970&2456504.84951&6&1.71&0.9942&0.05364&7662\\\hline
HD 211970&2456506.79608&0.47&1.86&0.9598&0.05385&6825\\\hline
HD 211970&2456506.80612&9.49&3.55&0.9492&0.05433&3374\\\hline
HD 211970&2456555.60902&3.43&1.15&0.729&0.05246&19106\\\hline
HD 211970&2456556.72102&-2.35&1.17&0.8122&0.05351&15289\\\hline
HD 211970&2456603.58858&6.93&1.03&0.576&0.05298&24254\\\hline
HD 211970&2456610.55366&-13.6&1.03&0.5361&0.05267&22999\\\hline
HD 211970&2456816.93005&-10.69&1.79&0.881&0.0543&7642\\\hline
HD 211970&2456818.87738&-4.42&1.08&0.7789&0.05211&25940\\\hline
HD 211970&2456866.72527&-5.89&1.05&0.7527&0.05193&27615\\\hline
HD 211970&2456871.73924&-2.41&1.13&0.7926&0.05235&16594\\\hline
HD 211970&2456876.83292&2.96&1.24&0.8458&0.05245&19119\\\hline
HD 211970&2456879.717&0.6&1.08&0.8207&0.05307&26605\\\hline
HD 211970&2457198.89566&-1.7&1.08&0.8817&0.05225&20889\\\hline
HD 211970&2457203.81436&2.47&1.15&0.9188&0.0538&23365\\\hline
HD 211970&2457260.7319&0&1.18&0.7327&0.05214&17234\\\hline
HD 211970&2457321.5985&-5.3&1.07&0.4997&0.05319&19970\\\hline
HD 211970&2457536.92111&8.42&1.24&0.9646&0.05503&15202\\\hline
HD 211970&2457555.8774&-2.5&1.05&0.8753&0.05301&24541\\\hline
HD 211970&2457614.77018&4.04&1.14&0.9375&0.05419&26397\\\hline
HD 211970&2457621.70646&-0.82&1.05&0.8482&0.05353&20522\\\hline
HD 211970&2458293.84928&4.73&1.07&-1&-1&8166\\\hline
HD 211970&2458329.78811&-3.74&0.96&-1&-1&7189\\\hline
HD 211970&2458355.67643&-5.71&0.87&-1&-1&6828\\\hline
HD 211970&2458357.60275&-5.85&1.04&-1&-1&4154\\\hline
HD 211970&2458357.60651&-5.93&0.85&-1&-1&4515\\\hline
\hline
HD 39855&2455255.7322&-2.96&1.03&0.2681&0.03949&17240\\\hline
HD 39855&2455200.69318&3.74&1.57&0.2162&-1&17422\\\hline
HD 39855&2455255.58202&-3.12&1.13&0.1975&0.03123&32336\\\hline
HD 39855&2455587.59977&1.07&1.26&0.1736&0.03222&20347\\\hline
HD 39855&2455587.60343&1.21&1.21&0.1694&0.033&20166\\\hline
HD 39855&2455669.49598&2.9&1.09&0.1835&0.03233&46028\\\hline
HD 39855&2455845.83806&0.78&1.31&0.1667&0.03186&21890\\\hline
HD 39855&2455852.86337&0&1.42&0.1635&0.03258&19112\\\hline
HD 39855&2455956.63819&-1.3&1.31&0.159&0.03182&36101\\\hline
HD 39855&2456281.72155&-2.11&1.37&0.1665&-1&23915\\\hline
HD 39855&2456284.67998&0.75&1.19&0.1669&0.03205&49974\\\hline
HD 39855&2456345.5798&8.65&1.22&0.0324&0&24537\\\hline
HD 39855&2456356.60133&-4.48&1.17&0.166&0.03208&31425\\\hline
HD 39855&2456358.57618&7.48&1.65&0.2344&0.03372&9358\\\hline
HD 39855&2456695.62685&-0.22&1.28&0.1807&0.03197&36301\\\hline
HD 39855&2456703.59108&1.3&1.13&0.1992&0.03261&31290\\\hline
HD 39855&2457029.69333&-3.62&1.25&0.1645&0.03183&53474\\\hline
HD 39855&2457053.61031&-0.06&1.3&0.1691&0.0319&40647\\\hline
HD 39855&2457123.51006&-1.78&1.1&0.1692&0.03184&33114\\\hline
HD 39855&2457267.88973&1.74&1.23&0.1609&0.03221&36153\\\hline
HD 39855&2457325.81241&-3.09&1.27&0.1599&0.03185&43264\\\hline
HD 39855&2457389.6711&-0.3&1.31&0.1606&0.03181&45327\\\hline
HD 39855&2457395.70227&0.32&1.5&0.1663&0.03228&17590\\\hline
HD 39855&2457448.5851&-3.36&1.11&0.1677&0.03151&53217\\\hline
HD 39855&2457471.52459&-2.9&1.31&0.1635&0.03179&30674\\\hline
\hline
HIP 35173&2455255.7322&-2.96&1.03&0.2681&0.03949&17240\\\hline
HIP 35173&2455200.71753&6.59&1.68&0.3799&-1&7620\\\hline
HIP 35173&2455252.62714&-5.32&1.19&0.2378&-1&18749\\\hline
HIP 35173&2455581.64495&-3.75&1.05&0.2705&0.04027&23076\\\hline
HIP 35173&2455956.68819&-3.17&1.02&0.2119&0.03946&24086\\\hline
HIP 35173&2455957.64998&-0.34&1.69&0.3684&0.04145&5933\\\hline
HIP 35173&2456284.73876&2.24&0.99&0.2187&0.03928&25479\\\hline
HIP 35173&2456291.77672&-4.32&0.91&0.2199&0.03931&21960\\\hline
HIP 35173&2456343.62157&-2.2&0.99&0.2358&0.03954&21343\\\hline
HIP 35173&2456354.57027&0.3&0.99&0.2249&0.03952&18295\\\hline
HIP 35173&2456357.61684&4.18&0.94&0.2159&0.03952&19493\\\hline
HIP 35173&2456610.82296&1.19&1.01&0.2045&0.03912&21836\\\hline
HIP 35173&2456692.67294&1.23&1&0.2117&0.03897&25266\\\hline
HIP 35173&2456696.62238&-0.12&1.06&0.2266&0.03968&19352\\\hline
HIP 35173&2456698.67353&1.45&0.85&0.2184&0.03961&22937\\\hline
HIP 35173&2456729.6363&3.45&0.96&0.1861&0.03963&25493\\\hline
HIP 35173&2456730.59016&3.84&1.04&0.2392&0.03957&19102\\\hline
HIP 35173&2456733.59673&3.92&0.93&0.2016&0.03951&23207\\\hline
HIP 35173&2457020.72449&5.8&0.93&0.2355&0.03905&25335\\\hline
HIP 35173&2457022.71547&5.23&0.88&0.2298&0.03988&26572\\\hline
HIP 35173&2457030.75951&0.67&0.99&0.2397&0.04002&23900\\\hline
HIP 35173&2457050.6369&-0.88&0.93&0.2361&0.03982&25437\\\hline
HIP 35173&2457053.622&-2.14&1.05&0.2335&0.04009&16408\\\hline
HIP 35173&2457060.6255&0.71&0.99&0.2171&0.03874&21750\\\hline
HIP 35173&2457069.61469&1.56&1.07&0.2294&0.03968&16376\\\hline
HIP 35173&2457118.54054&-0.58&0.93&0.2351&0.03967&26280\\\hline
HIP 35173&2457122.53949&-0.06&0.89&0.2355&0.03954&23231\\\hline
HIP 35173&2457324.87053&-3.88&1.42&0.3246&0.04145&9338\\\hline
HIP 35173&2457327.85226&1.26&1.13&0.2742&0.04025&14149\\\hline
HIP 35173&2457387.66829&-3.02&0.88&0.2645&0.04026&24824\\\hline
HIP 35173&2457396.65701&-0.68&1.04&0.2645&0.03957&19286\\\hline
HIP 35173&2457448.62086&-2.62&1.05&0.2613&0.04009&25338\\\hline
HIP 35173&2457471.54655&2.21&1.07&0.2645&0.04011&15211\\\hline
HIP 35173&2457499.53549&-2.28&0.87&0.2483&0.03952&30511\\\hline
HIP 35173&2457737.74108&-3.26&1.08&0.2444&0.039&26502\\\hline
HIP 35173&2457758.74084&1.8&0.9&0.2517&0.04008&22330\\\hline
HIP 35173&2457762.65955&0&1.07&0.2701&0.03987&23516\\\hline
HIP 35173&2457824.59241&-0.76&0.99&0.2622&0.03915&18743\\\hline
HIP 35173&2458469.83192&-0.13&1&-1&-1&4885\\\hline
\hline
HD 102843&2455255.7322&-2.96&1.03&0.2681&0.03949&17240\\\hline
HD 102843&2455200.83576&0&2.43&0.2557&-1&5708\\\hline
HD 102843&2455252.79165&-1.84&1.26&0.1589&-1&14798\\\hline
HD 102843&2455342.58244&-0.1&1.26&0.1976&0.03497&14379\\\hline
HD 102843&2455585.824&2.86&1.31&0.1616&0.03431&15311\\\hline
HD 102843&2455664.67782&-1.93&1.32&0.21&0.0337&15991\\\hline
HD 102843&2456093.53284&7.9&1.06&0.3384&0.03394&13810\\\hline
HD 102843&2456288.8541&4.36&1.15&0.1613&0.0343&19078\\\hline
HD 102843&2456345.79961&7.62&1.18&0.1877&0.03412&17302\\\hline
HD 102843&2456435.55492&4.65&1.22&0.1722&0.03345&19933\\\hline
HD 102843&2456694.81016&7.42&1.13&0.1816&0.03405&20088\\\hline
HD 102843&2456698.78069&5.44&1.17&0.1787&0.03419&18673\\\hline
HD 102843&2456701.73522&5.79&1.21&0.173&0.03419&17658\\\hline
HD 102843&2456730.72343&3.96&1.22&0.1926&0.03423&15830\\\hline
HD 102843&2456817.53906&6.99&1.37&0.2421&0.03412&14105\\\hline
HD 102843&2457023.84581&4.55&1.1&0.1748&0.03459&17749\\\hline
HD 102843&2457029.83073&3.92&1.08&0.1635&0.0334&20407\\\hline
HD 102843&2457065.81831&6.29&1.25&0.1613&0.03409&14638\\\hline
HD 102843&2457069.72688&4.24&1.22&0.1656&0.03448&14364\\\hline
HD 102843&2457119.70904&0.6&1.22&0.1677&0.03355&17706\\\hline
HD 102843&2457203.49054&-0.26&1.17&0.158&0.03388&18199\\\hline
HD 102843&2457206.51837&-0.76&1.1&0.1546&0.03372&20241\\\hline
HD 102843&2457390.84293&3.21&1.14&0.1521&0.03421&17726\\\hline
HD 102843&2457397.84159&-1.19&1.22&0.1608&0.03427&12809\\\hline
HD 102843&2457448.79186&-0.79&1.23&0.1565&0.03404&22385\\\hline
HD 102843&2457472.68105&-4.12&1.3&0.1531&0.03406&19522\\\hline
HD 102843&2457505.60482&-2.59&1.29&0.1656&0.03407&15096\\\hline
HD 102843&2457760.85002&-5.58&1.14&0.1479&0.0342&21243\\\hline
HD 102843&2457829.80843&-3.93&1.6&0.2068&0.0341&11045\\\hline
HD 102843&2458204.66791&-3.35&0.93&-1&-1&14073\\\hline
HD 102843&2458204.71316&-4.14&1&-1&-1&13154\\\hline
HD 102843&2458205.72943&-3.62&1.03&-1&-1&13330\\\hline
HD 102843&2458206.68341&-3.11&1.05&-1&-1&12668\\\hline
HD 102843&2458207.70655&-2.38&0.98&-1&-1&14300\\\hline
HD 102843&2458208.63571&-2.28&0.94&-1&-1&19760\\\hline
HD 102843&2458209.65381&-3.39&0.93&-1&-1&20254\\\hline
\hline
HD 103949&2455255.7322&-2.96&1.03&0.2681&0.03949&17240\\\hline
HD 103949&2455200.85365&1.87&1.57&0.2838&-1&7861\\\hline
HD 103949&2455252.80669&-4.49&1.19&0.2311&-1&18529\\\hline
HD 103949&2455342.59972&1.02&1.05&0.1979&0.04018&29944\\\hline
HD 103949&2455584.81107&1.87&1.03&0.2294&0.04103&23085\\\hline
HD 103949&2455664.70212&-1.29&1.1&0.2303&0.04016&24463\\\hline
HD 103949&2456093.54236&-0.55&0.99&0.2439&0.04117&16120\\\hline
HD 103949&2456284.85339&3.57&1&0.2749&0.0407&24711\\\hline
HD 103949&2456345.81626&-3.56&1.03&0.2768&0.0409&26003\\\hline
HD 103949&2456356.81978&-2.33&1.12&0.2896&0.04172&20181\\\hline
HD 103949&2456431.59876&-0.01&1.59&0.2572&0.04084&21637\\\hline
HD 103949&2456434.58574&-1.51&1.05&0.2804&0.04086&21799\\\hline
HD 103949&2456438.57712&0.27&1.13&0.2702&0.04101&19845\\\hline
HD 103949&2456501.4829&-2.1&1.24&0.2944&0.04111&14991\\\hline
HD 103949&2456693.77863&-4.52&0.98&0.2627&0.04093&33792\\\hline
HD 103949&2456696.74631&-2.93&1.04&0.2767&0.0402&22235\\\hline
HD 103949&2456700.78295&0.22&0.96&0.2344&0.04033&23764\\\hline
HD 103949&2456702.70173&-0.47&1.05&0.2669&0.04105&22201\\\hline
HD 103949&2456729.7549&-1.08&0.94&0.2666&0.03986&41798\\\hline
HD 103949&2456730.73207&0.41&1.02&0.2827&0.04008&24473\\\hline
HD 103949&2456816.56657&-1.35&1.28&0.2531&0.04054&23830\\\hline
HD 103949&2457022.81444&2.14&0.9&0.2581&0.04022&25317\\\hline
HD 103949&2457026.8595&-0.68&0.96&0.2575&0.04001&19568\\\hline
HD 103949&2457051.86844&-1.99&1&0.2518&0.0401&24487\\\hline
HD 103949&2457064.70854&-1.16&1.03&0.2449&0.04014&21884\\\hline
HD 103949&2457118.69169&-2.56&1.05&0.2416&0.04088&26524\\\hline
HD 103949&2457123.69519&0.34&1.05&0.2494&0.0411&21538\\\hline
HD 103949&2457200.48178&-1&1.09&0.2458&0.04072&23451\\\hline
HD 103949&2457389.80697&0.96&1.04&0.2319&0.04031&18497\\\hline
HD 103949&2457396.86666&2.12&0.97&0.2184&0.04079&21717\\\hline
HD 103949&2457450.78222&-0.99&1.1&0.2338&0.03996&23609\\\hline
HD 103949&2457472.71043&-1&1.04&0.2142&0.04002&23336\\\hline
HD 103949&2457478.70621&-0.58&0.99&0.2063&0.03953&37022\\\hline
HD 103949&2457505.62043&2.32&1.06&0.2133&0.03972&23868\\\hline
HD 103949&2457555.52785&-0.85&1.13&0.2765&0.04002&22711\\\hline
HD 103949&2457760.859&1.99&0.92&0.2064&0.04058&36733\\\hline
HD 103949&2457761.81112&3.34&0.91&0.2053&-1&36580\\\hline
HD 103949&2457765.79163&0&1.08&0.2084&0.04087&21426\\\hline
HD 103949&2457825.75799&1.52&0.95&0.2212&0.04043&26191\\\hline
HD 103949&2457862.66605&0.07&1.11&0.2366&0.03969&16008\\\hline
HD 103949&2458203.69321&1.08&0.87&-1&-1&21612\\\hline
HD 103949&2458204.75568&2.05&0.88&-1&-1&15853\\\hline
HD 103949&2458205.74082&2.23&0.88&-1&-1&22214\\\hline
HD 103949&2458206.67226&1.42&0.83&-1&-1&19816\\\hline
HD 103949&2458207.71827&1.48&0.81&-1&-1&21013\\\hline
HD 103949&2458208.62537&0.8&0.83&-1&-1&22948\\\hline
HD 103949&2458209.64388&0.82&0.79&-1&-1&20962\\\hline
HD 103949&2458264.57394&-0.63&0.91&-1&-1&19389\\\hline
\hline
HD 206255&2455255.7322&-2.96&1.03&0.2681&0.03949&17240\\\hline
HD 206255&2455427.71742&-2.66&1.3&0.1349&0.02973&21832\\\hline
HD 206255&2455439.73885&-4.03&1.32&0.1355&0.03028&18934\\\hline
HD 206255&2455796.7401&2.49&1.4&0.1427&0.03018&19100\\\hline
HD 206255&2455850.65044&2.32&1.26&0.1429&0.02996&19692\\\hline
HD 206255&2455850.65416&2.34&1.4&0.1444&0.02997&19669\\\hline
HD 206255&2456086.81919&-1.77&1.18&0.1315&0.02905&39826\\\hline
HD 206255&2456092.85872&-1.16&1.23&0.1297&0.03006&25348\\\hline
HD 206255&2456139.70543&1.73&1.27&0.1338&0.02995&23160\\\hline
HD 206255&2456144.7747&2.14&1.65&0.1404&0.03056&13672\\\hline
HD 206255&2456150.7277&-0.11&1.44&0.131&0.02972&18626\\\hline
HD 206255&2456504.83089&-3.38&1.32&0.1362&0.03007&20147\\\hline
HD 206255&2456506.78512&1.91&1.32&0.142&0.02958&20609\\\hline
HD 206255&2456550.62471&7.66&1.43&0.3088&0.02968&32231\\\hline
HD 206255&2456553.6186&4.55&1.47&0.1388&0.02961&21158\\\hline
HD 206255&2456604.54902&-4.1&1.23&0.1347&0.02951&26861\\\hline
HD 206255&2456817.86837&0.09&1.23&0.1319&0.02993&31069\\\hline
HD 206255&2456866.69534&-3.18&1.31&0.1284&0.02972&39119\\\hline
HD 206255&2456876.68366&-9.05&1.29&0.1275&0.02919&39943\\\hline
HD 206255&2457198.86731&-1.8&1.34&0.1267&0.02996&27749\\\hline
HD 206255&2457206.83583&-2.37&1.43&0.1242&0.02971&38134\\\hline
HD 206255&2457258.61859&-2.48&1.31&0.1298&0.02977&30047\\\hline
HD 206255&2457321.57596&5.79&1.37&0.1304&0.0296&26552\\\hline
HD 206255&2457327.61917&3.74&1.3&0.129&0.02982&24059\\\hline
HD 206255&2457536.90929&-5.04&1.31&0.1341&0.0296&23177\\\hline
HD 206255&2457555.86064&-5.31&1.21&0.1229&0.02888&43387\\\hline
HD 206255&2457614.71064&0&1.2&0.1293&0.02955&38830\\\hline
HD 206255&2457620.63101&-4.69&1.37&0.1404&-1&20427\\\hline
HD 206255&2458271.77631&2.13&1.19&-1&-1&17530\\\hline
HD 206255&2458293.81535&1.92&1.21&-1&-1&20410\\\hline
HD 206255&2458334.76763&-1.66&1.18&-1&-1&6918\\\hline
HD 206255&2458334.77142&0.67&1.21&-1&-1&7649\\\hline
HD 206255&2458354.67918&-2.19&1.44&-1&-1&4487\\\hline
HD 206255&2458354.68676&0&1.41&-1&-1&4517\\\hline
\hline
HD 21411&2455255.7322&-2.96&1.03&0.2681&0.03949&17240\\\hline
HD 21411&2455200.62078&5.04&1.34&0.2286&-1&13359\\\hline
HD 21411&2455430.89198&-0.25&1.36&0.2078&0.03409&20044\\\hline
HD 21411&2455584.57744&2.11&1.18&0.2027&0.03301&21664\\\hline
HD 21411&2455852.79817&-6.04&1.46&0.1842&0.03321&19016\\\hline
HD 21411&2456143.9235&19.35&1.93&0.1895&-1&12955\\\hline
HD 21411&2456281.63855&-2.43&1.33&0.2106&-1&25027\\\hline
HD 21411&2456290.59309&-3.22&0.92&0.1861&0.03291&39304\\\hline
HD 21411&2456343.53765&-1.08&1.04&0.1941&0.03295&24919\\\hline
HD 21411&2456358.52243&4.55&2.91&0.3238&0.03612&3820\\\hline
HD 21411&2456551.8202&-0.56&1.97&0.2034&-1&23523\\\hline
HD 21411&2456605.70767&-6.31&1.41&0.2024&0.03334&22103\\\hline
HD 21411&2456612.67223&3.34&1.25&0.1946&0.03323&30911\\\hline
HD 21411&2456697.5676&-4.21&1.04&0.1869&0.03265&32159\\\hline
HD 21411&2456866.90825&-2.66&1.32&0.1951&0.03274&26895\\\hline
HD 21411&2456882.92219&-1.71&1.34&0.1947&0.03284&36006\\\hline
HD 21411&2457022.63789&-3.14&1.01&0.196&0.03318&36914\\\hline
HD 21411&2457029.63113&3&0.96&0.1996&0.0331&30569\\\hline
HD 21411&2457053.52375&-1.31&1.2&0.1995&0.03282&21376\\\hline
HD 21411&2457061.56943&-1.68&1.05&0.1944&0.03258&23471\\\hline
HD 21411&2457260.8869&1.71&1.46&0.1854&0.03264&21551\\\hline
HD 21411&2457320.75509&-0.14&1.28&0.1889&0.0329&22339\\\hline
HD 21411&2457389.61809&0&0.98&0.1808&0.03256&35380\\\hline
HD 21411&2457622.92626&-0.93&1.29&0.1882&0.03299&25644\\\hline
HD 21411&2457741.62792&11.4&1.28&0.2188&0.03356&37117\\\hline
HD 21411&2457759.59214&12.12&0.98&0.4113&0.03345&26847\\\hline
HD 21411&2457765.62903&8.83&1.27&0.2267&0.03378&18377\\\hline
HD 21411&2458410.70071&0.12&1.15&-1&-1&6032\\\hline
HD 21411&2458410.70461&2.18&1.15&-1&-1&5966\\\hline
HD 21411&2458417.74652&17.64&1.39&-1&-1&3519\\\hline
HD 21411&2458417.75021&17.21&1.36&-1&-1&4550\\\hline
\hline
HD 64114&2455255.7322&-2.96&1.03&0.2681&0.03949&17240\\\hline
HD 64114&2455200.75154&-0.15&1.31&0.2264&-1&13346\\\hline
HD 64114&2455255.63455&-1.84&1.31&0.2038&0.03093&39062\\\hline
HD 64114&2455586.77045&-2.2&1.24&0.1658&0.03165&20608\\\hline
HD 64114&2455588.6773&-0.93&1.05&0.1662&0.03134&32884\\\hline
HD 64114&2455671.55327&-3.56&1.12&0.2606&0.03134&45495\\\hline
HD 64114&2456288.7264&6.35&1.06&0.1931&0.03201&41494\\\hline
HD 64114&2456290.78816&6.76&1.02&0.1901&0.03203&46531\\\hline
HD 64114&2456345.63885&-1.92&1.34&0.0318&0&27065\\\hline
HD 64114&2456695.68081&2.47&1.22&0.1977&0.03172&24676\\\hline
HD 64114&2456702.66536&-0.27&1.03&0.174&0.03123&42052\\\hline
HD 64114&2456733.60378&0.43&1.09&0.1725&0.0309&37678\\\hline
HD 64114&2456734.61888&0.66&1.03&0.1925&0.03118&41885\\\hline
HD 64114&2457021.71572&4.38&1&0.1766&0.03171&37198\\\hline
HD 64114&2457024.75524&4.53&1.05&0.1732&0.03228&42179\\\hline
HD 64114&2457053.66366&-1.5&1.38&0.179&0.03179&23599\\\hline
HD 64114&2457061.63104&0.62&1.12&0.1723&0.03153&23509\\\hline
HD 64114&2457066.61776&3.57&1.29&0.1754&0.03175&19556\\\hline
HD 64114&2457117.56268&1.4&1.12&0.257&0.03123&43344\\\hline
HD 64114&2457122.55954&1.09&1.14&0.1829&0.0312&43004\\\hline
HD 64114&2457387.7083&0.06&0.89&0.1676&-1&44362\\\hline
HD 64114&2457395.66312&-1.26&1.32&0.1695&0.03179&20384\\\hline
HD 64114&2457468.59249&-1.25&1.09&0.1994&0.03131&44770\\\hline
HD 64114&2457499.55607&-4.62&1.13&0.1788&0.03112&45070\\\hline
HD 64114&2457740.73137&0&1.2&0.1793&0.0316&34668\\\hline
HD 64114&2457761.70022&1.89&1.18&0.1765&-1&36217\\\hline
HD 64114&2457769.75899&-0.32&1.26&0.1745&0.03163&25866\\\hline
HD 64114&2457833.59771&-0.31&1.02&0.2028&0.03161&25333\\\hline
\hline
HD 8326&2455255.7322&-2.96&1.03&0.2681&0.03949&17240\\\hline
HD 8326&2455585.54494&1.02&0.93&0.2795&0.03895&23077\\\hline
HD 8326&2456141.9085&1.63&0.93&0.2996&0.03954&21201\\\hline
HD 8326&2456173.80187&-2.34&1.18&0.3144&0.03946&19470\\\hline
HD 8326&2456612.6198&7.2&1.15&0.2793&0.03956&20615\\\hline
HD 8326&2456866.87142&-3.69&0.97&0.3533&0.03979&24202\\\hline
HD 8326&2456877.86915&8.75&1.19&0.3872&0.03964&21502\\\hline
HD 8326&2457022.60447&2.08&0.78&0.3289&0.03911&24195\\\hline
HD 8326&2457261.87244&0&1.14&0.3374&0.04018&18113\\\hline
HD 8326&2457324.71174&-7.66&1.05&0.2769&0.03915&21787\\\hline
HD 8326&2457389.55141&12.24&0.85&0.3446&0.03958&22770\\\hline
HD 8326&2457617.86045&-2.06&0.95&0.2911&0.03929&22645\\\hline
HD 8326&2457739.57509&1.28&0.91&0.2429&0.03867&24241\\\hline
HD 8326&2457762.5492&-7.57&1.01&0.2622&0.0384&18459\\\hline
HD 8326&2457767.5642&-6.47&0.94&0.2491&0.03803&23125\\\hline
HD 8326&2458407.70844&-2.62&0.97&-1&-1&7325\\\hline
\hline
HIP 31609&2455255.7322&-2.96&1.03&0.2681&0.03949&17240\\\hline
HIP 31609&2458407.8526&-10.6&1.38&-1&-1&1890\\\hline
HIP 31609&2458408.85256&0.43&1.44&-1&-1&2290\\\hline
HIP 31609&2458409.84653&-0.03&1.61&-1&-1&1786\\\hline
HIP 31609&2458410.84883&5.98&1.39&-1&-1&1854\\\hline
HIP 31609&2458411.85426&6.32&1.21&-1&-1&2563\\\hline
HIP 31609&2458412.85539&7.35&1.37&-1&-1&2603\\\hline
HIP 31609&2458414.85554&6.52&1.68&-1&-1&1516\\\hline
HIP 31609&2458415.80466&0.65&1.52&-1&-1&2027\\\hline
HIP 31609&2458416.84913&-5.34&1.36&-1&-1&2229\\\hline
HIP 31609&2458417.78781&-4.53&1.8&-1&-1&1330\\\hline
HIP 31609&2458418.84782&3.99&1.37&-1&-1&2197\\\hline
HIP 31609&2458467.77358&-6.71&1.19&-1&-1&2795\\\hline
HIP 31609&2458467.8106&-4.61&1.51&-1&-1&2272\\\hline
HIP 31609&2458468.75877&-9.98&1.27&-1&-1&2637\\\hline
HIP 31609&2458468.81937&-12.14&1.21&-1&-1&2516\\\hline
HIP 31609&2458469.76285&-7.79&1.35&-1&-1&1833\\\hline
HIP 31609&2458469.82074&-10.52&1.54&-1&-1&1670\\\hline
HIP 31609&2458471.76184&-7.42&1.41&-1&-1&2250\\\hline
HIP 31609&2458471.83237&-12.98&1.43&-1&-1&1684\\\hline
HIP 31609&2458473.76748&-5.48&1.13&-1&-1&2773\\\hline
HIP 31609&2458473.83554&0&1.34&-1&-1&1845\\\hline
HIP 31609&2458474.71356&0.35&1.26&-1&-1&2174\\\hline
HIP 31609&2458474.82237&0.53&1.57&-1&-1&1908\\\hline
HIP 31609&2458475.73745&6.98&1.38&-1&-1&2239\\\hline
HIP 31609&2458475.83545&10.06&1.28&-1&-1&2198\\\hline
HIP 31609&2458476.76185&11.97&1.45&-1&-1&2264\\\hline
HIP 31609&2458476.84509&9.29&1.28&-1&-1&2364\\\hline
\hline
HIP 54373&2455255.7322&-2.96&1.03&0.2681&0.03949&17240\\\hline
HIP 54373&2455200.81388&-0.01&3.3&1.1365&-1&3251\\\hline
HIP 54373&2455254.65524&-5.75&1.38&1.2993&-1&9665\\\hline
HIP 54373&2455341.55657&-1&1.44&1.0644&0.06204&8371\\\hline
HIP 54373&2455582.82444&-1.95&1.35&1.4122&0.0628&8352\\\hline
HIP 54373&2455668.64626&3.41&1.53&1.1943&0.06075&6572\\\hline
HIP 54373&2455672.54785&11.35&1.81&1.1771&0.0637&5709\\\hline
HIP 54373&2455954.79749&3.3&1.45&1.5138&0.06136&7566\\\hline
HIP 54373&2455955.79678&4.54&1.53&1.4293&0.06109&6743\\\hline
HIP 54373&2455957.82077&9.43&3.81&1.3056&0.06466&2210\\\hline
HIP 54373&2455958.80249&4.63&1.7&1.4975&0.06263&5216\\\hline
HIP 54373&2455959.80289&3.33&1.51&1.5661&0.06282&6459\\\hline
HIP 54373&2455960.80708&-0.08&1.57&1.4675&0.06325&5887\\\hline
HIP 54373&2456086.51118&-0.64&1.48&0.9156&0.05982&8747\\\hline
HIP 54373&2456088.528&3.92&1.79&0.876&0.06047&5976\\\hline
HIP 54373&2456092.50926&2.43&2.68&1.0897&0.06162&3342\\\hline
HIP 54373&2456093.49839&0.85&1.5&0.8899&0.06177&4120\\\hline
HIP 54373&2456282.84604&-5.35&1.51&1.4307&0.06256&7443\\\hline
HIP 54373&2456288.83793&-1.11&1.3&1.5356&0.06108&7256\\\hline
HIP 54373&2456292.84185&7.97&1.32&1.4246&0.06095&7182\\\hline
HIP 54373&2456344.72208&-3.99&1.68&1.4487&0.06154&6179\\\hline
HIP 54373&2456353.74857&14.91&1.58&1.0695&0.06313&6197\\\hline
HIP 54373&2456356.78344&-11.1&1.66&1.4895&0.06219&5956\\\hline
HIP 54373&2456428.55774&8.18&1.86&1.0544&0.06211&5817\\\hline
HIP 54373&2456433.55267&-13.61&1.77&1.0966&0.0624&6182\\\hline
HIP 54373&2456434.5606&-12.48&3.04&1.1327&0.06307&2929\\\hline
HIP 54373&2456438.56787&7.56&2.2&1.0119&0.06025&4376\\\hline
HIP 54373&2456693.75493&-4.85&1.33&1.4456&0.06084&8806\\\hline
HIP 54373&2456696.73138&3.04&1.41&1.5517&0.06225&7912\\\hline
HIP 54373&2456698.75589&-2.94&1.43&1.4152&0.06235&7353\\\hline
HIP 54373&2456701.71882&4.8&1.51&1.4904&0.06203&6851\\\hline
HIP 54373&2456734.73037&-0.39&1.36&1.3109&0.06347&8708\\\hline
HIP 54373&2456735.71095&-1.82&1.42&1.421&0.06248&6898\\\hline
HIP 54373&2456818.49057&4.89&1.67&0.8612&0.0608&5877\\\hline
HIP 54373&2457021.84001&1.8&1.15&1.6562&0.06457&10807\\\hline
HIP 54373&2457026.82576&-0.99&1.28&1.6873&0.06407&11485\\\hline
HIP 54373&2457051.75433&6.1&1.32&1.6401&0.06481&8960\\\hline
HIP 54373&2457062.72414&0&1.48&1.4429&0.0605&7730\\\hline
HIP 54373&2457120.68354&0&1.43&1.2376&0.06137&9716\\\hline
HIP 54373&2457203.47387&-1.62&1.5&1.1355&0.06207&6658\\\hline
HIP 54373&2457387.8144&-4.92&1.26&1.7353&0.06515&8091\\\hline
HIP 54373&2457397.80848&0.66&1.5&1.7506&0.06575&6829\\\hline
HIP 54373&2457448.74521&-1.84&1.5&1.5171&0.06162&8295\\\hline
HIP 54373&2457473.66083&1.81&1.52&1.4173&0.06234&8871\\\hline
HIP 54373&2457478.66936&5.97&1.3&1.2089&0.06035&9693\\\hline
HIP 54373&2457499.6386&-3.47&1.49&1.0454&0.06117&9651\\\hline
HIP 54373&2457559.48686&-0.38&1.26&1.0418&0.06163&11132\\\hline
HIP 54373&2457759.83236&4.75&1.32&1.4273&0.06249&10217\\\hline
HIP 54373&2457824.71916&-2.79&1.74&1.4785&0.06239&7142\\\hline
HIP 54373&2457848.57423&2.37&1.43&1.4862&0.0641&9027\\\hline
HIP 54373&2458264.55224&-5&1.08&-1&-1&6014\\\hline
\hline
HD 24085&2455255.7322&-2.96&1.03&0.2681&0.03949&17240\\\hline
HD 24085&2455582.60234&-1.46&1.08&0.1352&0.02901&23031\\\hline
HD 24085&2455582.60495&5.07&1.2&0.1546&0.02952&15946\\\hline
HD 24085&2456145.91788&0.84&1.9&0.1528&0.0293&16842\\\hline
HD 24085&2456290.6024&0&1.09&0.1333&0.02865&44667\\\hline
HD 24085&2456551.82884&8.13&2.39&0.1996&-1&23351\\\hline
HD 24085&2456606.70671&6.58&1.52&0.1481&-1&18894\\\hline
HD 24085&2456698.55207&3.42&1.24&0.1376&0.02833&31582\\\hline
HD 24085&2456871.92807&-0.14&1.72&0.1322&0.0281&34132\\\hline
HD 24085&2457022.6474&-0.19&1.11&0.1333&0.02843&46090\\\hline
HD 24085&2457029.63619&-1.5&1.01&0.4486&0.02856&34399\\\hline
HD 24085&2457053.53884&-1.5&1.2&0.1425&0.02804&21504\\\hline
HD 24085&2457066.53534&-1.9&1.32&0.1423&0.02915&18847\\\hline
HD 24085&2457262.87843&-0.19&1.31&0.1383&0.02889&25690\\\hline
HD 24085&2457319.79348&-6.56&1.31&0.1408&0.02874&23074\\\hline
HD 24085&2457389.61345&-2.7&1.07&0.1307&0.0284&40886\\\hline
HD 24085&2457450.51648&-3.32&1.54&0.2745&0.02785&36913\\\hline
HD 24085&2457621.92363&1.77&1.46&0.1484&0.02901&19681\\\hline
HD 24085&2457741.63286&1.73&1.24&0.1384&0.02812&39691\\\hline
HD 24085&2457761.60444&0.11&0.99&0.1469&-1&36598\\\hline
HD 24085&2457766.5965&10.82&1.15&0.1383&0.02767&31505\\\hline
HD 24085&2458354.90321&0.8&1.44&-1&-1&5261\\\hline
HD 24085&2458410.70939&2.52&1.36&-1&-1&6274\\\hline
HD 24085&2458410.71321&0.55&1.36&-1&-1&5865\\\hline
HD 24085&2458417.75784&-1.6&1.22&-1&-1&9685\\\hline
\hline
HIP 71135&2455255.7322&-2.96&1.03&0.2681&0.03949&17240\\\hline
HIP 71135&2455342.66134&-0.84&1.81&0.6097&0.05082&6564\\\hline
HIP 71135&2455423.50522&2.75&1.56&0.4679&0.05022&6871\\\hline
HIP 71135&2455437.47257&5.23&1.44&0.4713&0.04998&7527\\\hline
HIP 71135&2455785.52424&5.55&1.71&0.6408&0.05044&5343\\\hline
HIP 71135&2455787.50123&4.29&1.58&0.5024&0.05042&6782\\\hline
HIP 71135&2455793.49665&0&2.74&0.9498&0.05275&3198\\\hline
HIP 71135&2455796.52697&1.33&2.21&0.6912&0.05062&3804\\\hline
HIP 71135&2455801.50064&-0.41&2.6&0.6665&0.05317&3256\\\hline
HIP 71135&2455802.48336&-1.87&2.52&0.8015&0.05235&3236\\\hline
HIP 71135&2455802.49089&2.99&4.08&0.6045&0.05228&2347\\\hline
HIP 71135&2455803.49751&5.78&2.06&0.6305&0.05074&4124\\\hline
HIP 71135&2455804.51336&6.2&2.46&0.8024&0.05122&3173\\\hline
HIP 71135&2455958.86941&7.28&1.7&0.4351&0.04957&5545\\\hline
HIP 71135&2456085.68322&0.21&1.99&0.5965&0.05015&5509\\\hline
HIP 71135&2456094.5975&0.37&2.22&0.6373&0.05039&4013\\\hline
HIP 71135&2456344.81656&-6.35&1.6&0.6323&0.04945&7260\\\hline
HIP 71135&2456347.86443&-4.46&1.47&0.5509&0.04972&6454\\\hline
HIP 71135&2456352.84684&-3.15&1.56&0.4161&0.04915&7537\\\hline
HIP 71135&2456355.82826&-1.64&1.62&0.473&0.04994&5795\\\hline
HIP 71135&2456357.83352&0.88&2.26&0.7276&0.05083&3488\\\hline
HIP 71135&2456433.68066&-9.55&1.96&0.5104&0.0495&5171\\\hline
HIP 71135&2456435.64872&-3.37&1.79&0.502&0.04961&7090\\\hline
HIP 71135&2456505.53671&-1.55&1.89&0.5166&0.05004&5215\\\hline
HIP 71135&2456505.54627&-2.26&1.8&0.5029&0.05041&4960\\\hline
HIP 71135&2456693.83985&-3.02&1.65&0.6546&0.04985&7995\\\hline
HIP 71135&2456696.85337&-4.25&1.66&0.4446&0.04991&7582\\\hline
HIP 71135&2456698.88682&1.17&2.02&0.5201&0.05064&4152\\\hline
HIP 71135&2456729.81559&-1.66&1.47&0.4786&0.04997&7362\\\hline
HIP 71135&2456733.82888&3.99&1.76&0.4666&0.04992&6573\\\hline
HIP 71135&2456734.82708&-2.76&1.64&0.5633&0.05004&6805\\\hline
HIP 71135&2456817.66181&-3.13&2.28&0.7852&0.05095&4477\\\hline
HIP 71135&2457067.85988&1.33&1.67&0.482&0.04926&6457\\\hline
HIP 71135&2457120.74646&-3.59&1.71&0.5834&0.04968&8725\\\hline
HIP 71135&2457204.58631&2.37&1.65&0.4521&0.05029&7820\\\hline
HIP 71135&2457265.48913&1.58&1.5&0.5086&0.05034&7955\\\hline
HIP 71135&2457449.88976&1.41&1.51&0.4015&0.04978&11195\\\hline
HIP 71135&2457500.82204&-6.79&1.69&0.5521&0.04979&8611\\\hline
HIP 71135&2457534.66035&4.6&1.92&0.7042&0.05033&5762\\\hline
HIP 71135&2457557.60976&-9.73&1.77&0.452&0.04921&8042\\\hline
HIP 71135&2457626.49196&4&1.52&0.5047&0.04976&7897\\\hline
HIP 71135&2457825.8283&-5.28&1.76&0.5633&0.04923&8850\\\hline
HIP 71135&2457850.80439&-4.94&1.82&0.4754&0.05018&6916\\\hline
HIP 71135&2458264.68979&-3.07&1.46&-1&-1&4145\\\hline
\enddata
\end{deluxetable*}

\section*{Acknowledgements}
This work has made use of data from the European Space Agency (ESA)
mission {\it Gaia} (\url{https://www.cosmos.esa.int/gaia}), processed by the
{\it Gaia} Data Processing and Analysis Consortium (DPAC,
\url{https://www.cosmos.esa.int/web/gaia/dpac/consortium}). Funding for the
DPAC has been provided by national institutions, in particular the
institutions participating in the {\it Gaia} Multilateral
Agreement. Support for this work was provided by NASA through Hubble
Fellowship grant HST-HF2-51399.001 awarded by the Space Telescope
Science Institute, which is operated by the Association of
Universities for Research in Astronomy, Inc., for NASA, under contract
NAS5-26555. The authors acknowledge the years of technical support from LCO staff in the successful operation of PFS, enabling the collection of the data presented in this paper.
\software{R package magicaxis \citep{robotham16},
fields\citep{fields}, minpack.lm \citep{elzhov16}.}

\bibliographystyle{aasjournal}
\bibliography{nm}  

\begin{thebibliography}{}
\expandafter\ifx\csname natexlab\endcsname\relax\def\natexlab#1{#1}\fi
\providecommand{\url}[1]{\href{#1}{#1}}
\providecommand{\dodoi}[1]{doi:~\href{http://doi.org/#1}{\nolinkurl{#1}}}
\providecommand{\doeprint}[1]{\href{http://ascl.net/#1}{\nolinkurl{http://ascl.net/#1}}}
\providecommand{\doarXiv}[1]{\href{https://arxiv.org/abs/#1}{\nolinkurl{https://arxiv.org/abs/#1}}}

\bibitem[{Anglada-Escud{\'e} {et~al.}(2013)Anglada-Escud{\'e}, Tuomi, Gerlach,
  Barnes, Heller, Jenkins, Wende, Vogt, Butler, Reiners, {et~al.}}]{anglada13}
Anglada-Escud{\'e}, G., Tuomi, M., Gerlach, E., {et~al.} 2013, Astronomy \&
  Astrophysics, 556, A126

\bibitem[{Anglada-Escud{\'e} {et~al.}(2016)Anglada-Escud{\'e}, Amado, Barnes,
  Berdinas, Butler, Coleman, de~la Cueva, Dreizler, Endl, Giesers,
  {et~al.}}]{anglada16}
Anglada-Escud{\'e}, G., Amado, P.~J., Barnes, J., {et~al.} 2016, Nature, 536,
  437

\bibitem[{{Astudillo-Defru} {et~al.}(2017){Astudillo-Defru}, {Forveille},
  {Bonfils}, {S{\'e}gransan}, {Bouchy}, {Delfosse}, {Lovis}, {Mayor}, {Murgas},
  {Pepe}, {Santos}, {Udry}, \& {W{\"u}nsche}}]{astudillo-defru17}
{Astudillo-Defru}, N., {Forveille}, T., {Bonfils}, X., {et~al.} 2017, \aap,
  602, A88, \dodoi{10.1051/0004-6361/201630153}

\bibitem[{Benedict {et~al.}(2016)Benedict, Henry, Franz, McArthur, Wasserman,
  Jao, Cargile, Dieterich, Bradley, Nelan, {et~al.}}]{benedict16}
Benedict, G., Henry, T., Franz, O., {et~al.} 2016, The Astronomical Journal,
  152, 141

\bibitem[{{Butler} {et~al.}(1996){Butler}, {Marcy}, {Williams}, {McCarthy},
  {Dosanjh}, \& {Vogt}}]{butler96}
{Butler}, R.~P., {Marcy}, G.~W., {Williams}, E., {et~al.} 1996, \pasp, 108,
  500, \dodoi{10.1086/133755}

\bibitem[{{Crane} {et~al.}(2010){Crane}, {Shectman}, {Butler}, {Thompson},
  {Birk}, {Jones}, \& {Burley}}]{crane10}
{Crane}, J.~D., {Shectman}, S.~A., {Butler}, R.~P., {et~al.} 2010, in
  \procspie, Vol. 7735, Ground-based and Airborne Instrumentation for Astronomy
  III, 773553

\bibitem[{{Dumusque}(2016)}]{dumusque16a}
{Dumusque}, X. 2016, \aap, 593, A5, \dodoi{10.1051/0004-6361/201628672}

\bibitem[{{Dumusque}(2018)}]{dumusque18}
---. 2018, \aap, 620, A47, \dodoi{10.1051/0004-6361/201833795}

\bibitem[{{Eker} {et~al.}(2015){Eker}, {Soydugan}, {Soydugan}, {Bilir}, {Yaz
  G{\"o}k{\c c}e}, {Steer}, {T{\"u}ys{\"u}z}, {{\c S}eny{\"u}z}, \&
  {Demircan}}]{eker15}
{Eker}, Z., {Soydugan}, F., {Soydugan}, E., {et~al.} 2015, \aj, 149, 131,
  \dodoi{10.1088/0004-6256/149/4/131}

\bibitem[{Elzhov {et~al.}(2016)Elzhov, Mullen, Spiess, Bolker, Mullen, \&
  Suggests}]{elzhov16}
Elzhov, T.~V., Mullen, K.~M., Spiess, A.-N., {et~al.} 2016

\bibitem[{{Feng} {et~al.}(2017{\natexlab{a}}){Feng}, {Tuomi}, \&
  {Jones}}]{feng17b}
{Feng}, F., {Tuomi}, M., \& {Jones}, H.~R.~A. 2017{\natexlab{a}}, \mnras, 470,
  4794, \dodoi{10.1093/mnras/stx1126}

\bibitem[{{Feng} {et~al.}(2016){Feng}, {Tuomi}, {Jones}, {Butler}, \&
  {Vogt}}]{feng16}
{Feng}, F., {Tuomi}, M., {Jones}, H.~R.~A., {Butler}, R.~P., \& {Vogt}, S.
  2016, \mnras, 461, 2440, \dodoi{10.1093/mnras/stw1478}

\bibitem[{{Feng} {et~al.}(2017{\natexlab{b}}){Feng}, {Tuomi}, {Jones}, {Xoo},
  {Barnes}, {Anglada-Escud{\'e}}, {Vogt}, \& {Butler}}]{feng17a}
{Feng}, F., {Tuomi}, M., {Jones}, H.~R.~A., {et~al.} 2017{\natexlab{b}}, AJ,
  submitted

\bibitem[{{Fischer} {et~al.}(2016){Fischer}, {Anglada-Escude}, {Arriagada},
  {Baluev}, {Bean}, {Bouchy}, {Buchhave}, {Carroll}, {Chakraborty}, {Dawson},
  {Diddams}, {Dumusque}, {Eastman}, {Endl}, {Figueira}, {Ford},
  {Foreman-Mackey}, {Fournier}, {Furesz}, {Gaudi}, {Gregory}, {Grundahl},
  {Hatsyzes}, {Hebrard}, {Herrero}, {Hogg}, {Howard}, {Johnson}, {Jorden},
  {Jurgenson}, {Latham}, {Laughlin}, {Loredo}, {Lovis}, {Mahadevan},
  {McCracken}, {Pepe}, {Perez}, {Phillips}, {Plavchan}, {Prato}, {Quirrenbach},
  {Reiners}, {Robertson}, {Santos}, {Sawyer}, {Segransan}, {Sozzetti},
  {Steinmetz}, {Szentgyorgyi}, {Udry}, {Valenti}, {Wang}, {Wittenmyer}, \&
  {Wright}}]{fischer16}
{Fischer}, D., {Anglada-Escude}, G., {Arriagada}, P., {et~al.} 2016, ArXiv
  e-prints.
\newblock \doarXiv{1602.07939}

\bibitem[{{Gaia Collaboration} {et~al.}(2018){Gaia Collaboration}, {Brown},
  {Vallenari}, {Prusti}, {de Bruijne}, {Babusiaux}, \&
  {Bailer-Jones}}]{gaiaDR2}
{Gaia Collaboration}, {Brown}, A.~G.~A., {Vallenari}, A., {et~al.} 2018, ArXiv
  e-prints.
\newblock \doarXiv{1804.09365}

\bibitem[{Gillon {et~al.}(2016)Gillon, Jehin, Lederer, Delrez, de~Wit,
  Burdanov, Van~Grootel, Burgasser, Triaud, Opitom, {et~al.}}]{gillon16}
Gillon, M., Jehin, E., Lederer, S.~M., {et~al.} 2016, Nature, 533, 221

\bibitem[{{Gregory}(2011)}]{gregory11}
{Gregory}, P.~C. 2011, \mnras, 410, 94,
  \dodoi{10.1111/j.1365-2966.2010.17428.x}

\bibitem[{Haario {et~al.}(2006)Haario, Laine, Mira, \& Saksman}]{haario06}
Haario, H., Laine, M., Mira, A., \& Saksman, E. 2006, Statistics and Computing,
  16, 339

\bibitem[{Haywood {et~al.}(2014)Haywood, Collier~Cameron, Queloz, Barros,
  Deleuil, Fares, Gillon, Lanza, Lovis, Moutou, {et~al.}}]{haywood14}
Haywood, R., Collier~Cameron, A., Queloz, D., {et~al.} 2014, Monthly notices of
  the royal astronomical society, 443, 2517

\bibitem[{{Kane} {et~al.}(2012){Kane}, {Ciardi}, {Gelino}, \& {von
  Braun}}]{kane12}
{Kane}, S.~R., {Ciardi}, D.~R., {Gelino}, D.~M., \& {von Braun}, K. 2012,
  \mnras, 425, 757, \dodoi{10.1111/j.1365-2966.2012.21627.x}

\bibitem[{Kass \& Raftery(1995)}]{kass95}
Kass, R.~E., \& Raftery, A.~E. 1995, Journal of the american statistical
  association, 90, 773

\bibitem[{{Kipping}(2013)}]{kipping13}
{Kipping}, D.~M. 2013, \mnras, 434, L51, \dodoi{10.1093/mnrasl/slt075}

\bibitem[{{Kopparapu} {et~al.}(2014){Kopparapu}, {Ramirez}, {SchottelKotte},
  {Kasting}, {Domagal-Goldman}, \& {Eymet}}]{kopparapu14}
{Kopparapu}, R.~K., {Ramirez}, R.~M., {SchottelKotte}, J., {et~al.} 2014,
  \apjl, 787, L29, \dodoi{10.1088/2041-8205/787/2/L29}

\bibitem[{Lomb(1976)}]{lomb76}
Lomb, N.~R. 1976, Astrophysics and space science, 39, 447

\bibitem[{Luger {et~al.}(2017)Luger, Sestovic, Kruse, Grimm, Demory, Agol,
  Bolmont, Fabrycky, Fernandes, Van~Grootel, {et~al.}}]{luger17}
Luger, R., Sestovic, M., Kruse, E., {et~al.} 2017, Nature Astronomy, 1, 0129

\bibitem[{{Malkov}(2007)}]{malkov07}
{Malkov}, O.~Y. 2007, \mnras, 382, 1073,
  \dodoi{10.1111/j.1365-2966.2007.12086.x}

\bibitem[{{Marcy} \& {Butler}(1992)}]{marcy92}
{Marcy}, G.~W., \& {Butler}, R.~P. 1992, \pasp, 104, 270,
  \dodoi{10.1086/132989}

\bibitem[{{Mayor} {et~al.}(2014){Mayor}, {Lovis}, \& {Santos}}]{mayor14}
{Mayor}, M., {Lovis}, C., \& {Santos}, N.~C. 2014, \nat, 513, 328,
  \dodoi{10.1038/nature13780}

\bibitem[{Nychka {et~al.}(2018)Nychka, Furrer, Paige, Sain, \& Nychka}]{fields}
Nychka, D., Furrer, R., Paige, J., Sain, S., \& Nychka, M.~D. 2018

\bibitem[{Pepe {et~al.}(2002)Pepe, Mayor, Rupprecht, Avila, Ballester, Beckers,
  Benz, Bertaux, Bouchy, Buzzoni, {et~al.}}]{pepe02}
Pepe, F., Mayor, M., Rupprecht, G., {et~al.} 2002, The Messenger, 110, 9

\bibitem[{{Pepe} {et~al.}(2010){Pepe}, {Cristiani}, {Rebolo Lopez}, {Santos},
  {Amorim}, {Avila}, {Benz}, {Bonifacio}, {Cabral}, {Carvas}, {Cirami},
  {Coelho}, {Comari}, {Coretti}, {De Caprio}, {Dekker}, {Delabre}, {Di
  Marcantonio}, {D'Odorico}, {Fleury}, {Garc{\'{\i}}a}, {Herreros Linares},
  {Hughes}, {Iwert}, {Lima}, {Lizon}, {Lo Curto}, {Lovis}, {Manescau},
  {Martins}, {M{\'e}gevand}, {Moitinho}, {Molaro}, {Monteiro}, {Monteiro},
  {Pasquini}, {Mordasini}, {Queloz}, {Rasilla}, {Rebord{\~a}o}, {Santana
  Tschudi}, {Santin}, {Sosnowska}, {Span{\`o}}, {Tenegi}, {Udry}, {Vanzella},
  {Viel}, {Zapatero Osorio}, \& {Zerbi}}]{pepe10}
{Pepe}, F.~A., {Cristiani}, S., {Rebolo Lopez}, R., {et~al.} 2010, in
  \procspie, Vol. 7735, Ground-based and Airborne Instrumentation for Astronomy
  III, 77350F

\bibitem[{Raftery(1995)}]{raftery95}
Raftery, A.~E. 1995, Sociological methodology, 25, 111

\bibitem[{{Rajpaul} {et~al.}(2015){Rajpaul}, {Aigrain}, {Osborne}, {Reece}, \&
  {Roberts}}]{rajpaul15}
{Rajpaul}, V., {Aigrain}, S., {Osborne}, M.~A., {Reece}, S., \& {Roberts}, S.
  2015, \mnras, 452, 2269, \dodoi{10.1093/mnras/stv1428}

\bibitem[{{Ribas} {et~al.}(2018){Ribas}, {Tuomi}, {Reiners}, {Butler},
  {Morales}, {Perger}, {Dreizler}, {Rodr{\'\i}guez-L{\'o}pez}, {Gonz{\'a}lez
  Hern{\'a}ndez}, {Rosich}, {Feng}, {Trifonov}, {Vogt}, {Caballero}, {Hatzes},
  {Herrero}, {Jeffers}, {Lafarga}, {Murgas}, {Nelson}, {Rodr{\'\i}guez},
  {Strachan}, {Tal-Or}, {Teske}, {Toledo-Padr{\'o}n}, {Zechmeister},
  {Quirrenbach}, {Amado}, {Azzaro}, {B{\'e}jar}, {Barnes}, {Berdi{\~n}as},
  {Burt}, {Coleman}, {Cort{\'e}s-Contreras}, {Crane}, {Engle}, {Guinan},
  {Haswell}, {Henning}, {Holden}, {Jenkins}, {Jones}, {Kaminski}, {Kiraga},
  {K{\"u}rster}, {Lee}, {L{\'o}pez-Gonz{\'a}lez}, {Montes}, {Morin}, {Ofir},
  {Pall{\'e}}, {Rebolo}, {Reffert}, {Schweitzer}, {Seifert}, {Shectman},
  {Staab}, {Street}, {Su{\'a}rez Mascare{\~n}o}, {Tsapras}, {Wang}, \&
  {Anglada-Escud{\'e}}}]{ribas18}
{Ribas}, I., {Tuomi}, M., {Reiners}, A., {et~al.} 2018, \nat, 563, 365,
  \dodoi{10.1038/s41586-018-0677-y}

\bibitem[{Ricker {et~al.}(2014)Ricker, Winn, Vanderspek, Latham, Bakos, Bean,
  Berta-Thompson, Brown, Buchhave, Butler, {et~al.}}]{ricker14}
Ricker, G.~R., Winn, J.~N., Vanderspek, R., {et~al.} 2014, Journal of
  Astronomical Telescopes, Instruments, and Systems, 1, 014003

\bibitem[{{Robotham}(2016)}]{robotham16}
{Robotham}, A.~S.~G. 2016, {magicaxis: Pretty scientific plotting with
  minor-tick and log minor-tick support}, Astrophysics Source Code Library.
\newblock \doeprint{1604.004}

\bibitem[{Scargle(1982)}]{scargle82}
Scargle, J.~D. 1982, The Astrophysical Journal, 263, 835

\bibitem[{{Schwab} {et~al.}(2016){Schwab}, {Rakich}, {Gong}, {Mahadevan},
  {Halverson}, {Roy}, {Terrien}, {Robertson}, {Hearty}, {Levi}, {Monson},
  {Wright}, {McElwain}, {Bender}, {Blake}, {St{\"u}rmer}, {Gurevich},
  {Chakraborty}, \& {Ramsey}}]{schwab16}
{Schwab}, C., {Rakich}, A., {Gong}, Q., {et~al.} 2016, in \procspie, Vol. 9908,
  Ground-based and Airborne Instrumentation for Astronomy VI, 99087H

\bibitem[{{Teske} {et~al.}(2018){Teske}, {Wang}, {Wolfgang}, {Dai}, {Shectman},
  {Butler}, {Crane}, \& {Thompson}}]{teske18}
{Teske}, J.~K., {Wang}, S., {Wolfgang}, A., {et~al.} 2018, \aj, 155, 148,
  \dodoi{10.3847/1538-3881/aaab56}

\bibitem[{Tuomi \& Anglada-Escud{\'e}(2013)}]{tuomi13}
Tuomi, M., \& Anglada-Escud{\'e}, G. 2013, Astronomy \& Astrophysics, 556, A111

\bibitem[{{Tuomi} {et~al.}(2013){Tuomi}, {Jones}, {Jenkins}, {Tinney},
  {Butler}, {Vogt}, {Barnes}, {Wittenmyer}, {O'Toole}, {Horner}, {Bailey},
  {Carter}, {Wright}, {Salter}, \& {Pinfield}}]{tuomi12}
{Tuomi}, M., {Jones}, H.~R.~A., {Jenkins}, J.~S., {et~al.} 2013, \aap, 551,
  A79, \dodoi{10.1051/0004-6361/201220509}

\bibitem[{{Van Eylen} {et~al.}(2019){Van Eylen}, {Albrecht}, {Huang},
  {MacDonald}, {Dawson}, {Cai}, {Foreman-Mackey}, {Lundkvist}, {Silva Aguirre},
  {Snellen}, \& {Winn}}]{VanEylen18}
{Van Eylen}, V., {Albrecht}, S., {Huang}, X., {et~al.} 2019, \aj, 157, 61,
  \dodoi{10.3847/1538-3881/aaf22f}

\bibitem[{{Vogt} {et~al.}(2014){Vogt}, {Radovan}, {Kibrick}, {Butler},
  {Alcott}, {Allen}, {Arriagada}, {Bolte}, {Burt}, {Cabak}, {Chloros},
  {Cowley}, {Deich}, {Dupraw}, {Earthman}, {Epps}, {Faber}, {Fischer}, {Gates},
  {Hilyard}, {Holden}, {Johnston}, {Keiser}, {Kanto}, {Katsuki}, {Laiterman},
  {Lanclos}, {Laughlin}, {Lewis}, {Lockwood}, {Lynam}, {Marcy}, {McLean},
  {Miller}, {Misch}, {Peck}, {Pfister}, {Phillips}, {Rivera}, {Sandford},
  {Saylor}, {Stover}, {Thompson}, {Walp}, {Ward}, {Wareham}, {Wei}, \&
  {Wright}}]{vogt14}
{Vogt}, S.~S., {Radovan}, M., {Kibrick}, R., {et~al.} 2014, Publications of the
  Astronomical Society of the Pacific, 126, 359, \dodoi{10.1086/676120}

\bibitem[{{Wenger} {et~al.}(2000){Wenger}, {Ochsenbein}, {Egret}, {Dubois},
  {Bonnarel}, {Borde}, {Genova}, {Jasniewicz}, {Lalo{\"e}}, {Lesteven}, \&
  {Monier}}]{wenger00}
{Wenger}, M., {Ochsenbein}, F., {Egret}, D., {et~al.} 2000, \aaps, 143, 9,
  \dodoi{10.1051/aas:2000332}

\end{thebibliography}
\end{document}